\documentclass[12pt]{article}

\usepackage{amssymb}
\usepackage{amsmath}
\usepackage{amscd}
\usepackage{latexsym}
\usepackage{graphicx}

\usepackage{enumerate}

\usepackage{cite}

\newcommand{\be}{\begin{equation}}
\newcommand{\ee}{\end{equation}}
\newcommand{\Dlt}{\Delta}
\newcommand{\dlt}{\delta}

\newcommand{\bt}{\beta}

\newcommand{\ep}{\varepsilon}
\newcommand{\al}{\alpha}
\newcommand{\ra}{\rightarrow}

\begin{document}

\begin{center}
{\Large {\bf Dynamic probabilistic decision networks} \\ [5mm]

V.I. Yukalov$^{1,2}$ and E.P. Yukalova$^3$ } \\ [3mm]

{\it $^1$Bogolubov Laboratory of Theoretical Physics, \\
Joint Institute for Nuclear Research, Dubna 141980, Russia \\ [2mm]

$^2$Instituto de Fisica de S\~ao Carlos, Universidade de S\~ao Paulo, \\
CP 369, S\~ao Carlos 13560-970, S\~ao Paulo, Brazil \\ [2mm]

$^3$Laboratory of Information Technologies, \\
Joint Institute for Nuclear Research, Dubna 141980, Russia } \\ [3mm]

{\bf E-mails}: yukalov@theor.jinr.ru, yukalova@theor.jinr.ru

\end{center}

\vskip 2cm

\begin{abstract}
A new type of decision networks is suggested and its operation is analyzed. The network 
nodes are represented by intelligent agents who can denote either some biological beings, 
like humans, or neurons of the brain, or the nodes of artificial intelligence. The specifics 
of the network are in the following: It is {\it probabilistic} in the sense that the choice, 
accomplished by each agent, is characterized by the related probability. It is {\it dynamic}, 
with the probabilities varying in time due to the exchange of information between the agents. 
It is {\it affective}, because the agents choose between alternatives by taking account of 
utility as well as of biases and emotions. In general, it is {\it heterogeneous}, being 
composed of the groups of agents with different properties, for instance having long-term 
memory and short-term memory. The network dynamics, caused by the information exchange, results 
in decision error decrease. The network operation is illustrated by the example starting with 
the Allais paradox, its resolution, and the decision error diminution in the process of 
decision dynamics with information exchange. Resorting to machine-learning techniques it 
is possible to regulate the behavior of the network agents forcing them to choose particular 
alternatives. 
     
\end{abstract}

\vskip 1cm
{\parindent=0pt
{\bf Keywords}: Decision network, Probabilistic choice, Decision dynamics, Collective 
decisions, Allais paradox}

\newpage

\section{Introduction}

Taking decisions, choosing between several alternatives (actions), is a tool that is in the 
core of numerous applications, including the operation of different networks and artificial 
intelligence \cite{Nilsson_1,Luger_2,Rich_3,Neapolitan_4,Russel_5,Poole_6}. For the 
mathematical description of a decision-network operation it is necessary to overcome several
problems. First, considering intelligent agents, it is desirable to take into account that
their decisions are based not merely of rational grounds evaluating the utility of alternatives, 
but also on biases and emotions. Second, several agents, connected in a network, in the process 
of decision making exchange information, so that the process becomes dynamic. Thus a realistic
decision network needs to combine two principal features, individual decision making taking
account of biases and emotions, as well as collective effects caused by information exchange
in the process of dynamic formation of decisions.    
 
The predominant theory describing individual behavior under uncertainty is nowadays the 
expected utility theory axiomatized by von Neumann and Morgenstern \cite{Neumann_7} and 
integrated with the theory of subjective probability by Savage \cite{Savage_8}. These 
normative theories define an optimal choice as that corresponding to the maximal utility. 
The maximization of utility is a rational, logical action with the deterministic prescription 
of choice. However, it is well known that each rational choice is accompanied by emotions
and biases that bring irrational flavor into the process of decision making. Qualitatively, 
the role of emotions in decision making has been discussed in numerous publications, as can 
be inferred from the review articles \cite{Jones_9,Scherer_10,Lerner_11,Scherer_12,Yukalov_13}, 
where further references can be found. 

However, mathematically incorporating affective features into a decision-making model is 
a rather complicated task that till now has not been very successful. In that route, there 
exist several problems: One has to define how to measure the influence of emotions on 
decision making. When considering multi-agent systems, it is necessary to take into account 
that emotions change with time due to the exchange of information between the agents. 
 
It is the main goal of the present paper to develop and analyze a decision network that
could effectively take account of affective features of the network agents and of the 
temporal evolution of the agent's decisions caused by the exchange of information. This 
model could be not merely useful for predicting the collective behavior of humans and other
biological networks, but it is also a necessary step for the development of approaches 
describing the functioning of neuronal networks and human-like artificial intelligence. 

The role of emotions in the functioning of intelligence, whether human or artificial, is 
very important. For quite a long time in history, the role of emotions in decision making 
has been underestimated as compared to the rational evaluation of utility. However, the 
modern point of view accepts that the role of emotions overall is very important. First 
of all, emotions do exist and it is practically impossible to get rid of them when making 
decisions. Certainly, sometimes emotions can be disruptive on rational decision making. 
But they can also be useful. Since in nature emotions have been formed during many years 
of evolution, they should have been appeared for some reason that should be useful for 
alive beings \cite{Darwin_14}. The following points are generally accepted as defining the
main helpful characteristics of emotions. 

(i) Emotions are generated by subconscious mechanisms that are much faster than conscious 
rational deliberation, hence they allow for grasping the main important facts before conscious 
logical conclusions. This can be crucial in many situations, even saving life of a decision 
maker. 

(ii) Emotions are useful in communication between alive beings, since they are the source of 
additional information allowing for a better understanding of situation and for the more 
appropriate reaction. 

(iii) Emotions help to choose an optimal alternative among the given set, especially under 
complicated tasks or under uncertainty. This is because emotions broaden the range of 
admissible actions, make more colorful the characteristics of alternatives, and direct the 
decision maker attention to the most relevant features.
                      
Till now, the standard way in the mathematical formulation of affective decision making has 
been similar to that used for taking account of behavioral features, when one extends the 
expected-utility theory to the so-called non-expected utility theory \cite{Machina_15,Machina_16} 
by modifying the expected utility to a generalized value functional, with adding some more 
terms supposedly characterizing irrational effects, as is done in the dual-process theory 
\cite{Yaari_17,Sun_18,Paivio_19,Evans_20,Stanovich_21,Kahneman_22}. In the same way, one 
composes a value functional, also called objective function, by adding more terms to expected
utility and assuming that these additional terms can take care of emotions 
\cite{Mellers_23,Bracha_24,Liu_25,Liu_26}. The other way is to try to find out the empirical
relations between feelings and gains or losses \cite{Charpentier_27}. 

The models, based on the modification of expected utility to some value functionals, have two 
weak points. First, all of them are purely descriptive, containing several parameters needing 
to be fitted, so that predictive power of such models is rather low. Second, and the most 
important, these models are based on deterministic grounds assuming that the aim of a decision 
maker is the maximization of the constructed value functional.       

However, the choice of real decision makers is not deterministic, but rather probabilistic.
All observation of human behavior, in experiments as well as in real life, suggest that 
preferences change in a random manner making the process of choice stochastic \cite{Carvajal_28}.
It is well known that even the same decision maker, when deciding on the same set of 
alternatives at different times, as a rule, often changes the chosen preferences. Such 
variations in decisions are unavoidable even for a single decision maker in the process of 
each decision. This is due to the fact that, as has been discovered in neurophysiological 
studies, the decision process in the brain is always accompanied by noise leading to random 
choice, the randomness being caused by generic variability and local instability of neural 
networks \cite{Werner_29,Arieli_30,Gold_31,Glimcher_32,Naqvi_33,Shadlen_34,Webb_35,Kurtz_36}. 
The variability arises not merely from limitations in the data available to the decision maker, 
but the choice itself is stochastic due to fundamental stochasticity in the physiological 
process of making decisions. Neurophysiological studies provide evidence that the process of 
decision making in the human brain is highly variable. Woodford \cite{Woodford_37} summarizes 
the modern point of view that considers randomness as an internal feature of functioning of 
the human brain. Thus decision making is a principally probabilistic process and has to be 
characterized rather by probabilities than merely by prescribed utility or value functionals. 
Then the central point, required for developing a probabilistic approach, is to formulate how 
the probabilities are to be defined.

In past years, there has been a number of claims that cognition is a quantum process, hence
needs to be described by the techniques of quantum theory. This trend, due to achievements of
quantum technology, has been quite fashionable and a huge flood of speculations on that topic
has been produced. An exhaustive list of the related literature can be found in the recent 
book by Wendt \cite{Wendt_38}. The basic hope in resorting to quantum techniques has been
to exploit the appearance in quantum probabilities of the interference effect, whose 
characteristics could be fitted for explaining some of the behavioral effects in decision 
making. The interference could be ascribed either to alternatives \cite{Wendt_38}, or to the 
modes of intrinsic noise \cite{Yukalov_39,Yukalov_40}, or to inconclusive events 
\cite{Yukalov_41}.

However, when the compared alternatives are classical macroscopic objects, and the agents 
are also classical macroscopic objects, quantum behavior of conscience or the brain seems 
to be an unjustified assumption. If the brain is a classical object, its functioning has to 
allow for classical description and no quantum theory is needed. If cognition would be quantum 
then the brain should also be a quantum object, which contradicts to all available 
neurophysiological information on the brain. Since the brain is classical, its functioning 
should be characterized by a classical theory. 

Moreover, the quantum description of cognition requires to introduce a number of quantum 
notions, such as Hilbert spaces, wave functions, operators of alternatives, statistical 
operators of decision makers, evolution operators, Hamiltonians etc. All these notions are 
hardly applicable to the functioning of the classical brain. In addition to numerous 
unnecessary notions, a large amount of the related parameters enters the description, with 
no information on where the values of these parameters should be taken from, except several 
that could be fitted from experiments for explaining the same experiments, which makes the 
fitting senseless. 
 
In the present paper, we suggest an approach, based on classical terms, characterizing a 
probabilistic dynamic decision network of intelligent agents. The approach enjoys the 
following features: It is {\it probabilistic}, with the choice of alternatives by each agent 
being characterized by probabilities. It is {\it dynamic}, with the choice probabilities 
varying with time because of information exchange. The nodes of the network can be represented 
by intelligent agents whose choice combines the rational evaluation of utility of alternatives 
as well as their emotional attractiveness. 
  
The suggested networks can characterize biological systems, such as human societies, or groups
of other biological beings experiencing emotions in the process of decision making, or it can 
describe neuronal networks with intrinsic variations. The method for characterizing the 
decision networks can find application to other types of dynamic probabilistic networks 
\cite{Mihajlovic_42,Friedman_43} and to complex agentic networks 
\cite{Brien_44,Lu_45,Hosseini_46,Sapkota_47}. 

It is necessary to stress that the approach, suggested in this paper, is principally different 
from the methods of machine learning. In the latter, as is known, a large set of data composing 
the input, is analyzed by means of statistical algorithms containing a number of parameters. 
The process of training implies that one finds what are to be the values of the parameters 
in order to obtain the desired result under given data. This is the learning stage. Then, 
employing the fitted parameters, the algorithms can perform tasks accomplishing actions under 
the similar circumstances. Briefly speaking, on the basis of the received data, a machine, 
playing the role of an agent, evaluates possible actions and chooses the optimal alternative 
with the largest utility factor. In that sense, a learning machine evaluates the utility 
factors of alternatives choosing the optimal one. 

What we consider is the next level, when there is not a single agent but a society of agents 
deciding between several alternatives whose utility factors have already been calculated. The 
agents interact with each other by exchanging information and make subsequent decisions. Thus 
we consider not a single decision process but a sequence of decisions of each agent in a society, 
under the given utility factors. In machine learning, the deciding machine takes a single 
decision under a fixed set of data. The difficulty is in a huge amount of data, requiring 
complicated algorithms for their evaluation. In our case, we deal not with a single machine 
making a single decision, but with a series of decisions of agents in a society of many agents, 
under unchanged data. In addition, we take account of emotions and imitation effects that have 
never been studied in machine learning.

We illustrate the approach by considering the Allais paradox, where its standard static form 
serves as an initial condition for the following dynamics. We demonstrate the role of emotions 
for the resolution of the paradox and we extend the consideration to the dynamics of decision 
making in a probabilistic decision network. We show that the deviation from the rational 
decision in the case of the Allais paradox diminishes with time due to the information 
exchange between the network agents.

\section{Decisions at initial stage}

We consider the situation where each agent of a society is confronted with the problem of 
choice from a set of several alternatives. At the initial moment of time, there has been yet 
no possibility for discussions between the society members, so that each agent makes a 
decision independently from others. A widespread way of characterizing a society is by 
associating its members with the nodes possessing a kind of spin whose different directions
represent different alternatives. This way is accepted in Ising-type social models
\cite{Indekeu_48,Castellano_49,Hooyberghs_50,Yukalov_51,Carlo_52}. Our approach is rather 
different. The decision network consists of agents who are characterized by the probabilities 
of choosing this or that alternative. The network operation consists of two parts that 
can conditionally be called static and dynamic. First, at the initial moment of time, 
when the society agents had yet no possibility of exchanging information with each other, 
they make a preliminary evaluation of utility and attractiveness of the alternatives 
resulting in a primary decision. Then the agents start exchanging information on their 
choices, and their decisions vary due to this information exchange. The first, static, step
of individual agents has been considered in Refs. \cite{Yukalov_53,Yukalov_54}. This primary 
evaluation of utility and attractiveness of alternatives is sketched in the present section, 
since it is necessary for being used in what follows as an initial condition for the subsequent 
dynamics.

\subsection{Behavioral probability}

Suppose an agent decides on the choice of an alternative from the set of alternatives
\be
\label{1}
 \mathbb{A} \; = \; \{ A_n: \; n = 1,2,\ldots, N_A \} \;  .
\ee
The probability of choosing an alternative $A_n$ is denoted as $p(A_n)$ satisfying the 
normalization 
\be
\label{2}
 \sum_{n=1}^{N_A} p(A_n) \;= \; 1 \; , \qquad 0 \; \leq \; p(A_n) \; \leq 1 \;  .
\ee
This quantity $p(A_n)$ is called {\it behavioral probability}, since it is assumed to 
characterize both the rational utility of the alternatives and the influence of 
decision-maker emotions associated with the considered alternatives. The behavioral 
probability involves two terms, one of which, $f(A_n)$, called utility factor, measures 
the rational usefulness of an alternative $A_n$ for the decision maker, while the other, 
$q(A_n)$, called attraction factor, quantifies the strength of the emotional attraction 
of the alternative $A_n$ for the decision maker. The form of the combination of $f(A_n)$ 
and $q(A_n)$ is based on the following foundation.

First of all, it is generally accepted that psychologically emotions and rational logic act 
additively to each other \cite{Frijda_1986,Lazarus_1991,Elster_1999,Roberts_2003,Evans_2004,
Deigh_2008,Deonna_2012}. Translating the psychological language to mathematics, this can be
interpreted literally as $f(A_n) + q(A_n)$.

Second, emotional processes and rational reasoning occur in different parts of the brain.
Emotional processes occur within the limbic system including amygdala, hippocampus, hypothalamus,
cingulate gyrus, and orbifrontal cortex, while rational reasoning is mainly located in the 
prefrontal cortex \cite{Adams_1985,Fuster_2008,Binder_2009}. Therefore, the processing of
emotions and reasoning occurs, in some sense additively to each other.   

Third, the dependence of the probability $p(A_n)$ on the terms $f(A_n)$ and $q(A_n)$ has to 
be such that the theory could reduce back to the classical rational decision theory when 
neglecting emotion effects, that is, the limiting condition
$$
p(A_n) \; \ra \; f(A_n) \; , \qquad q(A_n) \; \ra \; 0
$$ 
has to be valid.   

The simplest forms for the probability with respect to the terms $f(A_n)$ and $q(A_n)$, 
seems, could be multiplicative or additive forms. However, the multiplicative form does not
satisfy the above limiting condition. Hence, the simplest possible form is additive. In this 
way, the available arguments endorse the additive structure of the behavioral probability: 
\be
\label{3}
p(A_n) \; = \; f(A_n) + q(A_n) \;   .
\ee
 
The rational utility of an alternative $A_n$ is quantified by the {\it utility factor} 
$f(A_n)$ having the properties of the classical probability, in particular, satisfying 
the normalization condition
\be
\label{4}
  \sum_{n=1}^{N_A} f(A_n) \;= \; 1 \; , \qquad 0 \; \leq f(A_n) \; \leq 1 \;  
\ee
and the relation between the joint and conditional probabilities
\be
\label{5}
f(AB) \; = \; f(A|B) \; f(B) \qquad ( A,B \in \mathbb{A} ) \;   .
\ee
     
The {\it attraction factor} $q(A_n)$ characterizes the influence of emotions in the process 
of choice of the alternative $A_n$ showing how the considered alternative is attractive for
the decision maker. From properties (\ref{2}), (\ref{3}) and (\ref{4}), it follows that the 
attraction factor satisfies the normalization condition
\be
\label{6}
 \sum_{n=1}^{N_A} q(A_n) \;= \; 0 \;   ,
\ee
called the {\it alternation law}, and varies in the range
\be
\label{7}
 - f(A_n) \; \leq \; q(A_n) \; \leq 1 - f(A_n) \;  .
\ee

An alternative with the maximal utility factor $f(A_n)$ is termed the {\it most useful} and 
that one having the maximal attraction factor $q(A_n)$ is named the {\it most attractive}. 
The maximal behavioral probability $p(A_n)$ defines the {\it preferable alternative} or the
{\it optimal alternative},
\be
\label{8}
 p(A_{opt}) \; = \; \sup_n p(A_n) \;  .
\ee

\subsection{Utility factor}

Thus, each alternative is characterized by its utility and attractiveness. Utility is measured 
by a utility factor $f(A_n)$ evaluating the rational usefulness of the alternative for the 
decision maker. The utility factor is a function of the expected utility of an alternative, 
or of its value functional, or of its objective function denoted as $U(A_n)$. For concreteness,
we shall use the term expected utility, keeping in mind that, in general, this can be any value 
functional. 

The explicit expression for the utility factor can be defined by minimizing an information 
functional. The general information functional is given by the Kullback-Leibler form 
\cite{Kullback_55,Kullback_56} under the uniqueness condition
\be
\label{9}
\left| \; \sum_{n=1}^{N_A} f(A_n) \; U(A_n) \; \right| \; < \; \infty
\ee
required by the Shore-Johnson theorem \cite{Shore_57}. The Kullback-Leibler information 
functional, under the normalization condition (\ref{4}) and uniqueness condition (\ref{9}),
reads as
$$
 I[\; f\; ] \; = \; \sum_{n=1}^{N_A} f(A_n) \; \ln \; \frac{f(A_n)}{f_0(A_n)} +
\al \; \left[ \; 1 - \sum_{n=1}^{N_A} f(A_n) \; \right] +
$$
\be
\label{10}
+
\bt \; \left[ \; const - \sum_{n=1}^{N_A} f(A_n) \; U(A_n) \; \right] \;  ,
\ee
where $f_0(A_n)$ is a trial prior distribution, $\alpha$ and $\beta$ are the Lagrange 
multipliers. The multiplier $\alpha$ preserves the normalization condition (\ref{4}) and
$\beta$ plays the role of a belief parameter characterizing the decision maker belief in the 
fairness of the decision procedure \cite{Yukalov_53,Yukalov_54}. The trial distribution 
$f_0(A_n)$ can be accepted in the Luce \cite{Luce_58,Luce_59} form
\be
\label{11}
f_0(A_n)  \; = \; \frac{a_n}{\sum_{n=1}^{N_A} a_n} \qquad
( a_n \geq 0 ) \;  ,
\ee
in which $a_n$ is the attribute of an alternative $A_n$. Depending on whether the expected 
utilities (values) are positive or negative, the attributes are defined as
\begin{eqnarray}
\label{12}
a_n \; = \; \left\{ \begin{array}{ll}
U(A_n) \; ,             ~ & ~ U(A_n) \geq 0 \\
1/|U(A_n)| \; , ~ & ~ U(A_n) < 0
\end{array} \right. \; .
\end{eqnarray}

The minimization of the information functional (\ref{10}) results in the distribution
\be
\label{13}
f(A_n) \; = \; 
\frac{f_0(A_n) e^{\bt U(A_n)} }{\sum_{n=1}^{N_A} f_0(A_n) e^{\bt U(A_n)} } \; .
\ee
When a decision maker is neutral with respect to the fairness of the decision process, then
the belief parameter is zero ($\beta = 0$), so that the posterior and prior distributions 
coincide,
\be
\label{14}
 f(A_n) \; = \; f_0(A_n) \qquad ( \bt = 0 ) \;  .
\ee

\subsection{Attraction factor}

Attraction factor $q(A_n)$ quantifies emotions experienced by a decision maker in the 
process of choice. Generally, it is a random quantity lying in the interval (\ref{7}), which
is different for different agents and even for the same agent at different times. In principle,
it is possible to construct distribution functions over attraction factors in order to model
the behavior of particular agents in particular circumstances \cite{Kovalenko_60}. However,
when we need to predict the outcomes of decisions for unknown agents, we should be able to
make non-informative estimates of attraction factors. This can be done keeping in mind that 
for large groups of agents there should exist quantities characterizing the decision makers 
on average, that is on aggregate level. 

The definition of non-informative estimates for typical attraction factors can be done on
the basis of arithmetic averages. When a random quantity $y$ lies in the interval $[a,b]$,
its arithmetic average is $(a + b)/2$. If the interval boundaries are also random quantities
in the intervals $[a_1,a_2]$ and, respectively, $[b_1,b_2]$, then they also are estimated by 
means of arithmetic averages. Thus the non-informative estimate $\overline{y}$ for a random 
quantity $y$ is 
\be
\label{15}
\overline y \; = \; \frac{1}{2} \; ( \overline a + \overline b ) \;   ,
\ee
where
$$
\overline a \; = \; \frac{1}{2} \; ( a_1 + a_2) \;  , 
\qquad
\overline b \; = \; \frac{1}{2} \; ( b_1 + b_2) \; .
$$

The attraction factor for negative emotions, according to inequality (\ref{7}), is in the 
range 
\be
\label{16}
 - f(A_n) \; \leq \; q(A_n) \; < \; 0 \;  ,
\ee
while for positive emotions, it is in the range
\be
\label{17}
0 \; \leq  \; q(A_n) \; \leq \; 1 - f(A_n) \; .
\ee
The utility factor $f(A_n)$ is in the interval $[0,1]$, hence its non-informative estimate
is $1/2$. Therefore the non-informative estimates for the attraction factors are
\be
\label{18}
q_- \; = \; -\; \frac{1}{4} \; , \qquad q_+ \; = \; \frac{1}{4}  .
\ee

Estimating the attraction factors as $\pm 0.25$, we have to keep in mind the normalization 
conditions (\ref{2}), so that $f(A_n) + q(A_n)$ be always in the interval $[0,1]$.  
For a set of $N_A$ alternatives, the following proposition is valid.

\vskip 2mm
{\bf Theorem} (\cite{Yukalov_54}). Assume that there are $N_A$ alternatives $A_n$ arranged 
so that
\be
\label{19}
q(A_n) \; > \; q(A_{n+1}) \qquad ( n = 1,2,\ldots,N_A-1 ) \; ,
\ee
having the typical difference between the attraction factors
\be
\label{20}
 \Dlt \; = \; q(A_n) - q(A_{n+1}) \;  ,
\ee
whose average magnitude is defined by the non-informative estimate
\be
\label{21}
\overline q \; = \; 
\frac{1}{N_A} \sum_{n=1}^{N_A} |\; q(A_n) \; | \; = \; \frac{1}{4} \;  .
\ee
Then the attraction factors are
$$
q(A_n) \; = \; \frac{N_A - 2n +1}{2N_A} \qquad (N_A \; even) \; ,
$$
\be
\label{22}
q(A_n) \; = \; \frac{N_A(N_A - 2n +1)}{2(N_A^2-1)} \qquad (N_A \; odd) \; ,
\ee
depending on whether $N_A$ is even or odd.

\vskip 2mm

As examples, let us consider the cases with several alternatives. For a set of two 
alternatives, we have
\be
\label{23}
\{ \; q(A_n) : \; n = 1,2 \; \} \; = \; 
\left\{ \frac{1}{4} \; , ~ -\; \frac{1}{4} \right\} \;  ,
\ee
for three alternatives,
\be
\label{24}
\{ \; q(A_n) : \; n = 1,2,3 \; \} \; = \; 
\left\{ \frac{3}{8} \; , ~ 0 \; , ~ -\; \frac{3}{8} \right\} \;  ,
\ee
and for the case of four alternatives, we get
\be
\label{25}
\{ \; q(A_n) : \; n = 1,2,3,4 \; \} \; = \; 
\left\{ \frac{3}{8} \; , ~ \frac{1}{8} \; , 
~ -\; \frac{1}{8} \; , ~ -\; \frac{3}{8} \right\} \; .
\ee
  
Thus the magnitudes of attraction factors can be estimated as is explained above. In order 
to completely describe the considered alternatives, it is necessary to qualify their relative
attractiveness on the basis of a well defined criterion. Generally, there exist two types
of alternatives, those that can be explicitly formulated as mathematically defined lotteries
and those that are described in words. Respectively, the classification of the lotteries onto
more or less attractive is done differently for these different classes. 

Qualitatively described alternatives are classified as more or less attractive by comparing
their features from the point of view of characteristics generally accepted as positive or 
negative. For instance, such emotional inclinations as {\it uncertainty aversion} and 
{\it propensity for cooperation} are treated and felt as positive \cite{Perc_61,Jusup_62}. 

If the choice between alternatives is the choice between explicitly defined lotteries
\be
\label{26}
L_n \; = \; \{ \; x_i, p_n(x_i) : \; i = 1,2,\ldots \; \} \; ,
\ee
then it is possible to formulate a quantitative criterion for distinguishing the lotteries 
as more or less attractive \cite{Yukalov_54}. Let $u(x)$ be a utility function of a payoff $x$. 
The {\it lottery quality} is defined  as 
\be
\label{27}
 Q(L_n) \; \equiv \; \sum_i u(x_i) \; 30^{p_n(x_i)} \;   .
\ee
The derivation of the expression for the lottery quality $Q(L_n)$ is based on the findings 
in experimental neuroscience, which have discovered that, when making a choice, the main 
and foremost attention of decision makers is directed toward the payoff probabilities
\cite{Kim_63}, but not to payoffs. This implies that subjects evaluate higher the 
probabilities than the related payoffs. In mathematical terms, this can be formulated as the 
existence of different types of scaling for the quality of an alternative with respect to 
payoff utility and payoff probability, e.g., the payoff utility being scaled linearly, while 
the payoff probability, exponentially. Then, the following consideration explains the form 
of the lottery quality. Let us consider lottery (\ref{26}), whose expected utility consists
of the sum of the terms $u(x) p(x)$ corresponding to the quality terms $u(x) b^{p(x)}$. If
the payoff $x$ enjoys a high certainty \cite{Hillson_64,Hillson_65}, such that the 
probability $p(x) = 3/4$ or higher, the payoff $x$ is preferred. When the payoff is increased
by $\lambda > 1$ times, becoming $\lambda x$, while the payoff probability decreases to 
$p(x)/ \lambda$, the utility term $x p(x)$ does not change. But the quality term reads as
$\lambda u(x) b^{p(x)}/\lambda$. According to numerous empirical observation, increasing the 
payoff by an order, with $\lambda = 10$, makes the large payoff, preferable, although its
probability diminishes. The threshold, where the choice inverts, corresponds to the equality
$$
u(x) b^{p(x)} = \lambda u(x) b^{p(x)/\lambda}  \; ,
$$
from where
$$
b = \lambda^{\lambda /(\lambda-1)p(x)} \; .
$$
The latter, with $p(x) = 3/4$ and $\lambda = 10$, gives $b = 30$. 

A given set of lotteries $\{L_n\}$, can be characterized by their qualities $Q(L_n)$. Since 
the qualities are numbers, they can be explicitly compared. Between the lotteries $L_m$ and 
$L_n$ that is more attractive whose quality is higher. For instance, if $Q(L_m) > Q(L_n)$, 
then the attractiveness of $L_m$ is higher than that of $L_n$, so that
\be 
\label{28}
q(L_m) \; > \; q(L_n) \; ; \qquad 
Q(L_m) \; > \; Q(L_n) \;   .
\ee

It may happen that the lottery qualities are equal, $Q(L_m) = Q(L_n)$. Then one has to compare
the gain-loss numbers 
\be
\label{29}
N(L_n) \; \equiv \; N_+(L_n) - N_-(L_n)
\ee
that are the differences between the number of payoff gains $N_+(L_n)$ and payoff losses, 
$N_-(L_n)$, of the corresponding lotteries. If the qualities of two lotteries are equal, 
then that lottery is more attractive whose difference between the number of gains and losses 
in the set of payoffs $\{x_i\}$ of the corresponding lotteries is larger \cite{Yukalov_54}. 
Thus $L_m$ is more attractive then $L_n$, that is $q(L_m) > q(L_n)$, when
\be
\label{30}
{\rm either} ~~ Q(L_m) \; > \; Q(L_n) \; , ~~
{\rm or} ~~ Q(L_m) \; = \; Q(L_n) ~~ {\rm and}~~ N(L_m) \; > \; N(L_n) \; .
\ee
 
Estimating the utility and attraction factors, each agent obtains an estimate for the prior
behavioral probability. Being a member of a social system, the agent begins exchanging  
information with other members of the society, as a result of which the initial prior choice
can be corrected. As far as the evaluation of the given lottery quality is unambiguously 
prescribed, the comparison between the lotteries can be accomplished automatically by a
neuron network.

\subsection{Empirical justification}

Attraction factors of different individuals, of course, vary depending on variable individual 
features of decision makers, such as mood, biases, moral, culture, gender, age etc. 
Moreover, the attraction factor can vary even for the same decision maker at different times. 
In that sense, the attraction factor is a random variable taking the values in the range defined 
by condition (\ref{7}). However, as is known, random variables can enjoy well defined means. 
The quarter law for the attraction factor is exactly a non-informative mean, as is explained
in Sec. 2.3. Generally, there exist two types of possibilities. For each given pool of decision 
makers, one may search for an attraction factor distribution over the given group. The other
problem is to define the average attraction factor over the group. When one deals with a group
of people whose individual features are not known, the quarter law serves as a non-informative 
prior characterizing the average attraction factor. We show in the present section that a 
thorough analysis, accomplished over a large number of decision-making tasks, is in perfect 
agreement with the quarter law.         

In order to prove that our approach, taking into account emotion effects, is well justified 
by extensive experimental observations and enjoys at the aggregate level a higher predictive 
power than other classical decision-making models, let us consider empirical evidence. Let 
us start with a set of decision problems suggested in the classical paper by Kahneman and 
Tversky \cite{Kahneman_1979} who showed that in real life many decision tasks contradict
the classical expected decision theory. However, in our approach all such problems are easily 
solved without fitting parameters, if the non-informative prior for the attraction factor is 
used. Conversely, defining the attraction factors from experimental observations, the quarter 
law (\ref{21}) is obtained.   
      
The task consists in the choice between the lotteries $L_n = \{x_i, p_n(x_i) \}$, where $x_i$ 
is a payoff and $p_n(x_i)$, payoff probability. Accepting a linear utility function, the 
expected utility of the lottery $L_n$ is 
$$
 U(L_n) \; =\; \sum_i x_i p_n(x_i) \;  .
$$
Utility factors, in the case of objective unbiased choices, are calculated, according to 
Sec. 2.2, as
$$
f(L_n) \; =\; \frac{U(L_n)}{\sum_n U(L_n) } \qquad ( U(L_n) \geq 0 )
$$
for semi-positive expected utilities, and as
$$
f(L_n) \; =\; \frac{|\;U(L_n)\;|^{-1}}{\sum_n |\; U(L_n)\;|^{-1} } \qquad 
( U(L_n) < 0 )   
$$
for negative expected utilities. The lottery quality is given by (\ref{27}). A group of 
around $100$ participants had to chose the optimal lottery from the set of several lotteries. 
The typical statistical error was about $\pm 0.1$.

For each lottery, we can define the utility factor $f(L_n)$, lottery quality $Q(L_n)$, and   
evaluate the attraction factor as a non-informative prior $\pm 0.25$, thus deriving the 
predicted probability $p(L_n)$. Comparing with empirical data, we keep in mind that, in 
reality, attraction factors can vary for different choices with different lotteries. The 
observed experimental probabilities $p_{exp}(L_n)$ are defined as the fractions of 
participants choosing the related lottery. What is important is the choice at the aggregate 
level, when the final data are averaged over the choices. Thus, to check the quarter law, we 
need to calculate the average over the lotteries, as in (\ref{21}), and to compare 
$\overline{q}$ with the non-informative prior $0.25$. An example of calculations is given 
below. 

{\it Problem}. Choose the optimal between the lotteries
$$
L_1\; = \; \{ 2.5, ~ 0.33 \; | \; 2.4, ~ 0.66 \; | \; 0 , ~ 0.01 \} \; ,
\qquad
L_2 \; =\; \{ 2.4, ~ 1 \} \; .
$$

For the utility factors we have
$$
 f(L_1) \; = \; 0.501 \; , \qquad f(L_2) \; =\; 0.499 \;  ,
$$
so that, since $f(L_1) > f(L_2)$, according to the expected utility theory, one should prefer 
the lottery $L_1$. Contrary to this, the majority choose the lottery $L_2$. The latter is
in agreement with our theory, since the quality of the lottery $L_2$ is higher,
$$
 Q(L_1) \; = \; 30.3 \; , \qquad Q(L_2) \; =\; 72 \;    ,
$$
which implies that $q(L_1) < q(L_2)$, hence the non-informative priors for the attraction 
factors are 
$$
  q(L_1) \; = -\; 0.25 \; , \qquad q(L_2) \; =\; 0.25 \;   ,
$$
which gives the predicted behavioral probabilities
$$
p(L_1) \; = \; f(L_1) - 0.25 \; = \; 0.25 \; ,  \qquad
p(L_2) \; = \; f(L_2) + 0.25 \; = \; 0.75 \;  
$$
because of which the lottery $L_2$ is preferable. This prediction is in good agreement with 
the experimental data \cite{Kahneman_1979}
$$
p_{exp}(L_1) \; = \; 0.18 \; , \qquad   p_{exp}(L_2) \; = \; 0.82 \;    
$$
that, within the accuracy of the experiment, equal the predicted probabilities.
 
Conversely, it is possible to find the experimental attraction factor 
$$
q_{exp}(L_n) \; = \; p_{exp}(L_n) - f(L_n)  \; ,
$$
for each lottery, calculate the aggregate attraction factor
$$
q_{exp} \; \equiv \; \frac{1}{N_A} \sum_{n=1}^{N_A} |\; q_{exp}(L_n) \; |  ,
$$ 
and to compare it with the quarter law (\ref{21}).
    
Accomplishing the above-described calculations for all $18$ lotteries adduced by Kahneman 
and Tversky \cite{Kahneman_1979}, we find the experimental attraction factor $q_{exp} = 0.27$,
which is in perfect agreement with the quarter law giving $0.25$. Within the accuracy of the
experiment $0.1\%$, these results coincide with each other. Thus, taking into account emotions
allows us to make predictions, at the aggregate level, involving no fitting parameters. Let 
us emphasize that the choice of participants deciding on the considered lotteries cannot be
explained in the frame of the classical expected decision theory, where the attraction factor
is zero.    

We have also accomplished the described calculations for a large set of lotteries recently
studied in Ref. \cite{Murphy_2018}. The authors considered three classes of lotteries, $27$ 
lotteries containing only gains, $19$ lotteries with only losses, and $21$ mixed lotteries 
with both gains and losses. The same experiment was repeated after two weeks, with randomly 
changing the order of the lottery choices. Altogether, this makes $134$ decision tasks. In 
each class of the lotteries, the experimental aggregate attraction factor is found to be the 
same, $q_{exp} = 0.22$, which perfectly agrees with the quarter law. 

These results cannot be predicted by expected utility theory or other decision-making models,
discussed in the Introduction, without crippling expected decision theory by introducing 
specially designed value functionals and  fitting parameters. These models, deforming the 
expected utility theory, called non-expected utility theories \cite{Machina_15,Machina_16},
require the use of fitting parameters for each considered case and have no predictive power.

\section{Operation of decision network}

After the agents have formed their primary opinions, they start communicating with each other 
by exchanging information on their choices. They also can have the propensity of imitating 
each other \cite{Apesteguia_66}. As a result of the imitation and information exchange, the 
agents opinion can vary. In this section, we develop an approach describing dynamics in a 
network of intelligent agents.

\subsection{Probability dynamics}

Let us consider a network of $N$ agents, enumerated by the index $j = 1,2,\ldots,N$, 
deciding on a choice of an alternative from the set $\{A_n\}$ of $N_A$ alternatives 
labelled by $n = 1,2,\ldots,N_A$. An agent represents an individual decision maker, whose 
behavioral probability of choosing an alternative $A_n$ at time $t$ is $p_j(A_n,t)$. 
An agent can also represent a typical subject of a group of $N_j$ decision makers, 
when the behavioral probability has the meaning of a fraction of subjects from the group
$j$ choosing an alternative $A_n$ at time $t$,
\be
\label{31}     
 p_j(A_n,t) \; = \; \frac{N_j(A_n,t)}{N_j} \; , \qquad
N_j \; = \; \sum_{n=1}^{N_A} N_j (A_n,t) \;  .
\ee
An agent can represent as well a neuron of a neural network. Also, an agent can be a node 
in a network of artificial intelligence acting under the presence of intrinsic noise. The 
mathematics is the same for all mentioned representations, because of which a general 
representative of such networks will be called an agent. 

The behavioral probabilities of all agents are, of course, normalized,
\be
\label{32}
 \sum_{n=1}^{N_A} p_j(A_n,t) \; = \; 1 \; , \qquad
0 \; \leq \; p_j(A_n,t) \; \leq \; 1 \;  .
\ee
Similarly, the utility factors for all agents are also normalized,
\be
\label{33}
 \sum_{n=1}^{N_A} f_j(A_n,t) \; = \; 1 \; , \qquad
0 \; \leq \; f_j(A_n,t) \; \leq \; 1 \;   .
\ee
From the above normalization conditions, it follows that the attraction factors satisfy the 
alternation law
\be
\label{34}
  \sum_{n=1}^{N_A} q_j(A_n,t) \; = \; 0 \;  ,
\ee
being in the interval
\be
\label{35}
- f_j(A_n,t) \; \leq \; q_j(A_n,t) \; \leq 1 - f_j(A_n,t) \;   .
\ee

At the initial moment of time, the network is composed of separate agents described above, 
so that one returns back to the crowd of independent agents each of which is characterized 
by its initial behavioral probability, utility factor, and attraction factor,
\be
\label{36}
p_j(A_n) \; \equiv \; p_j(A_n,0) \; , 
\qquad 
f_j(A_n) \; \equiv \; f_j(A_n,0) \; ,
\qquad 
q_j(A_n) \; \equiv \; q_j(A_n,0) \; .
\ee
The prior value of the probability plays the role of an initial condition
\be
\label{37}
 p_j(A_n,0) \; = \; f_j(A_n) + q_j(A_n) \;  .
\ee

Decision makers, being the members of a society, have inclination to imitate the behavior 
of others \cite{Apesteguia_66}, which is termed {\it imitation} or {\it herd effect}. The 
strength of this effect will be characterized by the parameter $\varepsilon_j$ that can 
be varied in the range
\be
\label{38}
0 \; \leq \; \ep_j \; \leq 1 \qquad ( j = 1,2,\ldots,N) \; .
\ee

Dynamics of decision making by intelligent agents is described by the system of equations
$$
p_j(A_n,t+\tau) \; = \; ( 1 - \ep_j) \; 
[\; f_j(A_n,t) + q_j(A_n,t) \; ] +
$$
\be
\label{39}   
+
\frac{\ep_j}{N-1} \sum_{i(\neq j)}^N  [\; f_i(A_n,t) + q_i(A_n,t) \; ]  ,
\ee
where $\tau$ is a delay time and  $j = 1,2,\ldots,N$. Time can be measured in units of $\tau$,
so that in what follows time $t$ can be treated to be dimensionless taking the values 
$t = 1,2,\ldots$.   

The dependence on time of utility factors can be due to time discounting \cite{Frederick_67}. 
When the studied period of time is not long as compared with the typical discounting time, 
it is possible to keep utility factors as constants. Attraction factors vary with time 
according to the law 
\be
\label{40}
q_j(A_n,t) \; = \; q_j(A_n) \exp\{ - M_j(t) \} \;   ,
\ee
where $M_j(t)$ is the $j$-th agent memory based on the information received by the agent in 
the process of interactions with other members. At the beginning, there is no yet any memory, 
which implies
\be
\label{41}
 M_j(t)\; = \; 0 \qquad ( t < 1) \;   .
\ee
Memory appears after the first step and accumulates by the law
\be
\label{42}
M_j(t) = \Theta( t - 1) \; \sum_{t'=1}^t \; \sum_{i=1}^N J_{ji}(t,t') \mu_{ji}(t')  \; ,  
\ee
where $\Theta(t)$ is a unit-step function, $J_{ji}(t,t')$ is a function describing the 
strength of influence by agent $i$ on agent $j$, due to the interactions between the 
agents $i$ and $j$ during the time from $t'$ to $t$, under acquiring the information gain 
$\mu_{ji}(t')$ received by the agent $j$ from the agent $i$ at time $t'$. 

The information gain can be taken in the Kullback-Leibler form \cite{Kullback_55,Kullback_56} as
\be
\label{43}
 \mu_{ji}(t) \; = \; 
\sum_{n=1}^{N_A} p_j(A_n,t) \; \ln \; \frac{p_j(A_n,t)}{p_i(A_n,t) } \; .
\ee

In general, the interaction strength $J_{ij}$ can depend on the location of the interacting 
agents. Depending on the nature of the considered agents, there can happen two cases. One case
is when a network is composed not of intelligent alive beings, but of mechanistic nodes
rigidly fixed at spatial locations. Then the agent interactions depend on the distance between 
the agents, and one needs to specify the dependence of $J_{ij}$ on the indices. Therefore 
the structure of a network constitutes a graph, usually on a plane, composed of nodes (agents 
with a fixed prescribed location) and edges connecting the nodes \cite{Albert_68,Brooks_69}. 
The properties of such networks depend on their particular spatial structure. 

The other principally different case is when one considers a network formed by intelligent 
alive beings, such as humans or even animals. In human or animal societies the alive beings 
are not tied to fixed locations, but move and do not form any fixed graphs. Moreover, 
interactions in human societies do not depend on the distance between its agents who communicate 
at any distance by means of various messenger devices, such as phone, WhatsApp, Telegram, Teams,
Facebook, YouTube etc. The societies, where the agents are not bound by fixed locations,
and communicate at any distance, are characterized by the interaction strength independent of 
the spatial indices measuring the distance between the agents.

In the present paper, we consider the networks characterizing the societies of intelligent 
agents, such as human beings. These agents are not nailed to fixed spatial locations on a plane
and do not form any graphs, but they freely move and are able to communicate at arbitrary 
distance. Therefore the effective interactions $J_{ij}$ of the agents cannot depend on the 
distance between the agents. It is important to stress that this is not a mean-field 
approximation, but the exact consequence of the realistic nature of intelligent agents.  
 
In this way, for a society of intelligent agents, who can freely move and interact at 
arbitrary distance, the interaction strength has to be taken in the form
\be
\label{44}
 J_{ji}(t,t') \; = \; \frac{J_j(t,t')}{N-1} \;    .
\ee
This form has to be accepted when considering human or animal societies. 

In the case of neural networks, neuron interactions can depend on neuron locations. But, if 
their interactions are sufficiently long-range, the interaction function can be taken in 
form (\ref{44}). Brain studies have found \cite{Saito_70} that axons, connecting neurons, have 
lengths often exceeding the dimension of the neuronal cell body by several orders of magnitude. 
These extreme axonal lengths imply that neurons have mastered efficient mechanisms for long 
distance signaling. These elaborate mechanisms are required for neuronal functioning and 
maintenance of the nervous system. Neurons can fine-tune long distance signaling through 
calcium wave propagation and bidirectional transport of proteins, vesicles, and Messenger RNA 
along microtubules. Thus, the long-range interaction (\ref{44}) should be a good approximation
for neuronal networks.  
 
In this way, the memory can be presented in the form
\be
\label{45}
M_j(t) = \Theta( t - 1) \; \sum_{t'=1}^t \frac{J_j(t,t')}{N-1} \;
\sum_{i=1}^N \mu_{ji}(t')  \;   .
\ee

Generally, memory can keep information for different times. The ultimate types of memory are 
the long-term and short-term memories. Under long-term memory, the interactions practically 
do not depend on time, being constant,
\be
\label{46}
 J_j(t,t') \; = \; J_j \qquad (long-term) \; ,
\ee
while under short-term memory the interactions are local in time, when only the last step is 
remembered,
\be
\label{47}
 J_j(t,t') \; = \; J_j \dlt_{t t'} \qquad (short-term) \;  .
\ee
Thus the long-term memory can be written as
\be
\label{48}
 M_j(t) = \Theta( t - 1) \;  \frac{J_j}{N-1} \; \sum_{t'=1}^t \; 
\sum_{i=1}^N \mu_{ji}(t')  \qquad (long-term) \; ,    
\ee
and the short-term memory takes the form
\be
\label{49}
 M_j(t) = \Theta( t - 1) \;  \frac{J_j}{N-1} \;
\sum_{i=1}^N \mu_{ji}(t)  \qquad (short-term) \;   .
\ee
 
From the neuropsychological point of view, memory consists of several parts accomplishing 
different tasks. In our case, we are interested in the functioning of memory as a whole 
in the process of decision making. The same underlying neural systems that are critical 
for memory are going to be critical for decision-making. This decision-making memory is 
the faculty, which enables to encode, store, and retrieve obtained information over time 
in order to make decisions \cite{Squire_1987,Weber_2006,Redish_2015}. We never decide in 
isolation from past experiences, but rather our decisions are guided by prior knowledge 
and by valuations rooted in memory. In the process of making a choice, memory and emotions
are closely related and impact each other simultaneously. Due to the complexity of memory,
there is no one precise optimal choice, but our preferences are volatile and subject to 
change. This gives to decision theory additional justification for being probabilistic.     

Furthermore, it is clear that no memory can be perfect, but they are subject to fading,
distortion, and becoming inoperative \cite{Squire_1987,Weber_2006,Redish_2015}. Such a 
memory fading is, of course, different in different people. Some make decisions being 
based on their experience during very long past, and some quickly forget distant events
and judge on the basis of only recent experience. Thus, it is possible to separate 
population into groups according to their feature of possessing long-term (long-range) 
memory or short-term (short-range) memory depending on whether decisions are taken by
analysing long past history or remembering just recent events. It is in that sense we 
distinguish between the groups of population with short-term and long-term memory, keeping 
in mind the decision-making memory.   

The collective decision of a society with respect to an alternative $A_n$ is composed of the 
weighted sum of the group probabilities
$$
p(A_n,t) \; = \; \sum_{j=1}^N c_j(t) \; p_j(A_n,t) \; ,
$$
where $c_j(t) = N_j/N$ is a fraction of population in each group. This fraction, generally, 
can be either fixed or can depend on time. When there are no transitions from one group to
another, then $c_j$ does not depend on time. While, if the number of members in a group can 
vary, that is when there can occur transitions between the groups, then $c_j(t)$ is a function 
of time. 

In this way, the dynamics of a probabilistic choice, described by equations (\ref{39}),
takes into account the rational utility of alternatives, as well as their emotional 
attractiveness, the existence of agents with different type of memory, and the imitation 
effect. The utility factor can vary with time due to time discounting \cite{Frederick_67}. 
The attraction factor depends on time through the exchange of information with other members 
of the society. The overall decision process ends when the agents come to a firm conclusion 
on their choice among alternatives. This corresponds to finding a stable stationary state 
of the dynamic equations (\ref{39}), which would characterize a stable stationary state of 
the network. In some ultimate situations it may happen that the stationary state does not
exist, but agent's opinions experience permanent oscillations.

\subsection{Binary choice}

A very often case is when one considers the choice between two alternatives, say $A_1$ 
and $A_2$. In that case, it is possible to simplify the notation by setting
$$
p_j(A_1,t) \; \equiv \; p_j(t) \; , \qquad p_j(A_2,t) \; \equiv 1- p_j(t) \; ,
$$
$$
f_j(A_1,t) \; \equiv \; f_j(t) \; , \qquad f_j(A_2,t) \; \equiv 1- f_j(t) \;
$$
\be
\label{50}
q_j(A_1,t) \; = \;  q_j(t) \; , \qquad q_j(A_2,t) \; = \; - q_j(t) \;  .
\ee
Let us also consider the case of two groups of agents, when $j = 1,2$. Assume that the agents
in these groups are distinguished by different types of memory. Let the first group enjoy
the long-term memory 
\be
\label{51}
M_1(t) = \Theta( t - 1) \; \sum_{t'=1}^t \mu_{12}(t')  \qquad (long-term) \; ,
\ee
while the second group has the short-term memory 
\be
\label{52}
M_2(t) = \Theta( t - 1) \; \mu_{21}(t)  \qquad (short-term) \;  ,
\ee
where the interaction strength is set as $J_i=1$. The information gain (\ref{43}) takes 
the form
\be
\label{53}
 \mu_{ij}(t) = p_i(t) \ln \; \frac{p_i(t)}{p_j(t)} + 
[ \; 1 - p_i(t)\; ] \; \ln \; \frac{1-p_i(t)}{1-p_j(t)}  \;  .
\ee
Note that $\mu_{jj} = 0$.   

In what follows, we use the notation
\be
\label{54}
f_j \; \equiv \; f_j(0) \; = \; f_j(A_1,0) \; , \qquad
 q_j \; \equiv \; q_j(0) \; = \; q_j(A_1,0) \;  .
\ee
Keeping in mind the situation when the utility of alternatives does not change during the 
process of decision making, we have
\be
\label{55}
 f_j(t) \; = \; f_j(A_1,t) \; = \; f_j \;  .
\ee
The attraction factor becomes
\be
\label{56}
  q_j(t) \; = \; q_j \; \exp\{ - M_j(t) \; \} \;  .
\ee
In that way, we obtain the equations of dynamic affective decision making for the binary 
choice of two groups, 
\be
\label{57}
p_j(t + 1) = (1 - \ep_j) [\; f_j + q_j(t) \; ] + \ep_j [\; f_i + q_i(t) \; ] \;   ,
\ee
where $i \neq j$, with $i,j = 1,2$. The initial conditions are
\be
\label{58}
 p_j(0) \; = \; f_j + q_j \;  .
\ee

Depending on the treated situation, there can be the following cases. Both groups have the 
same type of memory, either long-range or short-range, with the difference between the 
groups in the initial conditions. This case, however, is not as interesting, as far as the 
group dynamics will be close to each other. Another, more interesting, possibility is when 
the groups differ by the type of their memory, while the initial conditions can be either 
the same or different.

\section{Allais paradox and its resolution}

To illustrate how the developed probabilistic approach of affective decision making works,
let us study the Allais paradox \cite{Allais_71}. First, we reconsider the standard setup,
when each agent makes a single-step individual decision. Then this primary step will serve 
as an initial condition for the following dynamics in a network of communicating agents. 
Even before the agents start exchanging information, the probabilistic approach shows that 
the paradox is, actually, explained by taking account of emotions. After the resolution of 
the Allais paradox by individual decision makers, the multi-step dynamics of opinions in a 
decision network will be treated in the following section.

In the Allais paradox, subjects make choices that are incompatible with expected utility 
theory as well as with each other. One chooses between the lotteries
\be
\label{59}
L_1 \; = \; \{\; 1, ~1 \; \} \; , \qquad 
L_2 \; = \; \{ \; 1, ~0.89 \; | \; 5, ~0.10 \; | \; 0, ~ 0.01 \; \} \; .
\ee
The majority of subjects choose the first lottery that is more certain, which, according to 
utility theory, means that $U(L_1) > U(L_2)$. Then one decides on the lotteries
\be
\label{60}
L_3 \; = \; \{\; 5,~ 0.1\; | \; 0, ~ 0.9 \; \} \; , \qquad 
L_4 \; = \; \{\; 1, ~0.11 \; | \; 0, ~0.89 \; \} \; .
\ee
The majority prefers $L_3$, where the payoff is larger, which implies that $U(L_3) > U(L_4)$. 

The lottery expected utilities are 
$$
U(L_1) \; = \; u(1) \; , \qquad 
U(L_2) \; = \; 0.01 u(0) + 0.89 u(1) + 0.1 u(5) \; ,
$$
\be
\label{61}
U(L_3) \; = \; 0.9 u(0) + 0.1 u(5) \; , \qquad
 U(L_4) \; = \; 0.89 u(0) + 0.11 u(1) \; ,
\ee
where $u(x)$ is an arbitrary utility function. From the inequality $U(L_1) > U(L_2)$,
it follows
$$
0.11 u(1) \; > \; 0.01 u(0) + 0.1 u(5) \; , 
$$
while from $U(L_3) > U(L_4)$, one has
$$
0.11 u(1) \; < \; 0.01 u(0) + 0.1 u(5) \; .
$$
The last two inequalities are in evident contradiction showing that the utility theory is
not obeyed. It is useful to stress that this paradox does not depend on the concrete utility
function $u(x)$. The paradox arises if one follows the standard expected utility theory.

However, the present probabilistic approach does not lead to any paradox. For simplicity, 
let us consider a neutral decision maker, with the belief parameter $\beta = 0$ and the 
linear utility function $u(x) \propto x$. Then for the utility factors, defined in 
(\ref{13}), we have
\be
\label{62}
 f(L_n) \; = \; \frac{U(L_n)}{\sum_n U(L_n) } \;  .
\ee

Let us compare the first two lotteries, $L_1$ and $L_2$. For the utility factors (\ref{62}), 
one has
\be
\label{63}     
f(L_1) \; = \; 0.418 \; , \qquad f(L_2) \; = \; 0.582 \; ,
\ee
so that the second lottery $L_2$ looks more useful. Nevertheless, the lottery qualities 
(\ref{27}) are
\be
\label{64}
Q(L_1) \; = \; 30 \; , \qquad Q(L_2) \; = \; 27.7 \;   ,
\ee
showing that the first lottery $L_1$ is more attractive. 

Estimating the attraction factors by the non-informative priors (\ref{18}), we have the 
lottery attraction factors $q(L_1) = 0.25$, while $q(L_2) = - 0.25$. This gives the 
behavioral probabilities 
$$
p(L_1) \; = \; f(L_1) + q(L_1) \; = \; 0.418 + 0.25 \; = \; 0.668 \; ,
$$
\be
\label{65}
p(L_2) \; = \; f(L_2) + q(L_2) \; = \; 0.582 - 0.25 \; = \; 0.332 \;   .
\ee
In this way, the first lottery $L_1$ turns out to be preferable, since $p(L_1) > p(L_2)$. 

Similarly, considering the choice between $L_3$ and $L_4$, we find the utility factors
\be
\label{66}
f(L_3) = 0.82 \; , \qquad f(L_4) = 0.18 \; 
\ee
and the lottery qualities
\be
\label{67}
Q_3 = 7.03 \; , \qquad Q_4 = 1.45 \; .
\ee
This tells us that $L_3$ is more useful as well as more attractive than $L_4$, hence 
certainly will be preferred, as far as $p(L_3) > p(L_4)$. 

In this way, the standard Allais paradox is the contradiction between the behavior of 
decision makers and the standard expected utility theory. However there is no any 
contradiction between the present probabilistic theory and the actual behavior of decision 
makers.

\section{Dynamics of Allais paradox}

After the individual decision makers have made the choice explaining the origin of the 
Allais paradox, they start communicating with each other exchanging information on their 
actions. How then their decisions will be changing with time? To answer this question, we 
need to consider the dynamics of decisions in the network, as is described in Sec. 3.  

Let the network consist of two groups of agents, one group of agents with long-term memory, 
and the other, with short-term memory. The agents decide on the choice between the lotteries
$L_1$ and $L_2$, given in (\ref{59}). This implies that we have the case of binary choice, 
as in Sec. 3.2, where we need to set $j = 1,2$. 
  
The dynamics of a social network is described by equations (\ref{57}) that for the 
considered case take the form
$$
p_1(t + 1) \; = \; (1 - \ep_1) [\; f_1 + q_1(t) \; ] + \ep_1 [\; f_2 + q_2(t) \; ] \; ,
$$
\be
\label{68}
p_2(t + 1) \; = \; (1 - \ep_2) [\; f_2 + q_2(t) \; ] + \ep_2 [\; f_1 + q_1(t) \; ] \;   .
\ee
Recall that $p_1(t)$ is the behavioral probability that a member of the first group, having 
long-term memory, chooses the first lottery $L_1$ at time $t$, while $p_2(t)$ is the 
behavioral probability that the member of the second group, possessing short-term memory,
chooses the first lottery $L_1$ at time $t$. 

The initial conditions are
\be
\label{69}
p_1(0) \; = \; f_1 + q_1 \; , \qquad p_2(0) \; = \; f_2 + q_2 \; .
\ee 
The attraction factors 
\be
\label{70}
q_1(t)\;  = \; q_1 \exp\{ - M_1(t) \} \; \qquad 
q_2(t) \; = \; q_2  \exp\{ - M_2(t) \} \;   ,
\ee
characterizing the emotions of the agents, depend on the type of memory. Since the first 
group is assumed to be composed of agents with long-term memory, while the second group, 
of agents with short-term memory, this implies that
\be
\label{71}
M_1(t) \; = \; \Theta( t - 1) \; \sum_{t'=1}^t \mu_{12}(t')  \; , \qquad
M_2(t) \; = \; \Theta( t - 1) \;  \mu_{21}(t) \;   ,
\ee
where the information gain $\mu_{ij}$ is given by (\ref{53}).  
  
Let us consider the period of time, when the utility factors, evaluating the usefulness of
the available lotteries, can be treated as constant objective quantities. In the considered
situation, they are
\be
\label{72}
f_1\; = \; f_2\; \equiv \; f=0.418 \; .
\ee
In the process of the Allais choice, the first lottery is more attractive, with the 
attraction factor that can be evaluated by the quarter law $0.25$. This situation is 
assumed to be accepted by the group with long-term memory. The agents with short-term 
memory can be more emotional than those with the long-term memory, so that their 
attraction factor is spread over the whole admissible region (\ref{35}). Thus, we analyze 
the opinion dynamics for the initial attraction factors
\be
\label{73}
 q_1 \; = \; 0.25 \; , \qquad - 0.418 \;  < \; q_2 \; < 0.582 \;  .
\ee
The imitation parameters of the agents are assumed to be close to each other, which allows us
to set $\varepsilon_1 = \varepsilon_2 \equiv \varepsilon$.  

When the network (society) is not homogeneous, the situation becomes more involved, and the 
decision dynamics depends on the network parameters. Since the initial utility factors are fixed
by (\ref{72}) and the attraction factor $q_1$ of the first group at the initial time is given
as $q_1 = 0.25$, there are the free parameters, $q_2$ in the range (\ref{73}) and the imitation 
parameters $\varepsilon_j$ in the range (\ref{38}).

First, we may notice that taking the difference of the equations in (\ref{68}), we have
\be
\label{74}
 p_1(t+1) - p_2(t+1) = [\; 1 - (\ep_1 + \ep_2) \; ] \; [\; f_1 + q_1(t) \; ] -
 [\; 1 - (\ep_1 + \ep_2) \; ] \; [\; f_2 + q_2(t)\; ] \;  .
\ee
From here, it is clear that if $\varepsilon_1 + \varepsilon_2 = 1$, then the probabilities
$p_1(t)$ and $p_2(t)$ become equal for all $t \geq 1$, even when at the initial moment of time
they were different. 
      
When the imitation effect is not essential, so that $\varepsilon$ is close to zero, then there 
exist the following dynamic regimes depending on the emotional characteristics of the group 
with short-term memory. Below we summarize the changes from the initial probability 
$p_i \equiv p_i(0) = f_i + q_i$ to the final probability $p_i^* = f_i + q_i^*$, where $q_i^*$
is the final attraction factor. 

\vskip 2mm
(i) {\it Negative emotions}. In the range of the initial attraction factor $q_2 \in (-0.418, 0)$, 
corresponding to the negative emotions of the group with short-term memory, the probabilities 
of choosing the first lottery tend with time $t \ra \infty$ to the values
\be
\label{75}
p_1(t) \; \ra \; f \; , \qquad p_2(t) \; \ra \; p_2^* \; < \; f \qquad
(-0.418<q_2<0 ) \;   ,
\ee
where $f = 0.418$. If $p_2(0)$ is close to zero, the decisions of the second group exhibit 
oscillations at the beginning of the decision making process. 

\vskip 2mm
(ii) {\it Neutral case}. When the decision makers of the second group are emotionally neutral, 
with $q_2 = 0$, then 
\be
\label{76}
p_1(t) \; \ra \; f \; , \qquad p_2(t) \; = \; f \qquad (q_2=0) \;   ,
\ee
where again $f = 0.418$.
  
\vskip 2mm
(iii) {\it Weak positive emotions}. In the case of moderately positive emotions in the second 
group, such that $q_2 \in (0, 0.25)$, the opinion dynamics tend to
\be
\label{77}
p_1(t) \; \ra \; p^* \; > \; f \; , \qquad p_2(t) \; \ra \; p^* \; > \; f
\qquad
(0 < q_2 < 0.25) \;   ,
\ee
with $p^*$ being a common consensual limit. Note that if $q_2 = 0.25$, then 
$p_1(t) = p_2(t) = 0.668$. 

\vskip 2mm
(iv) {\it Strong positive emotions}. When $q_2 \in (0.25, 0.582)$, then
\be
\label{78}
 p_1(t) \; \ra \; f \; , \qquad p_2(t) \; \ra \; p_2^* \; > \; f
\qquad
(0.25 < q_2 < 0.582) \;  .
\ee
If $p_2(0)$ is close to $1$, it exhibits oscillations at the beginning of the decision process.

\vskip 2mm

Different dynamic regimes of the choice by two groups exchanging information, without 
imitation, are illustrated in Figs. $1$ and $2$. As is seen, the decision makers with 
long-term memory usually tend to minimize the influence of emotions in the process of 
exchanging information, since the agents tend to evaluate the lotteries by their objective 
utility factor $f$. At the same time, the decision makers with short-term memory continue 
to be influenced by emotions, although the role of emotions also diminishes. Depending 
on whether their emotions are negative, neutral, or positive, the decision makers with 
short-term memory evaluate the first lottery differently, either even lower then its 
objective utility factor $f$, equally to $f$, or higher then $f$. Therefore, allowing to 
the agents to share with each other the information on their actions leads to the elimination 
of the Allais paradox for the agents with long-term memory. However, the agents with 
short-term memory continue to depend on their emotions, though less than at the initial time.  

As an illustration of diminishing emotion influence for agents exchanging information, let 
us summarize the results of Fig. 1 for the variation of the attraction factors from their 
initial values $q_i$ to the final values $q_i^*$, where $q_1$ and $q_1^*$ are the attraction 
factors for the agents with long-term memory, while $q_2$ and $q_2^*$ are the attraction 
factors for the agents with short-term memory:   
$$   
q_1 = 0.25 \longmapsto q_1^* = 0 ; \quad q_2 = -0.42  \longmapsto q_2^* = -0.32  \qquad  (Fig.~1a) \; ,
$$
$$
q_1 = 0.25 \longmapsto q_1^* = 0 ; \quad q_2 = -0.3   \longmapsto q_2^* = -0.26  \qquad  (Fig.~1b) \; ,
$$
$$
q_1 = 0.25 \longmapsto q_1^* = 0 ; \quad q_2 = -0.1   \longmapsto q_2^* = -0.098  \qquad (Fig.~1c) \; ,
$$
$$
q_1 = 0.25 \longmapsto q_1^* = 0 ; \quad q_2 = 0      \longmapsto q_2^* = 0  \qquad  (Fig.~1d) \; ,
$$
$$
q_1 = 0.25 \longmapsto q_1^* = 0.1; \quad q_2 = 0.1    \longmapsto q_2^* =  0.1  \qquad   (Fig.~1e) \; ,
$$
$$
q_1 = 0.25 \longmapsto q_1^* = 0 ; \quad q_2 = 0.26   \longmapsto q_2^* = 0.23  \qquad   (Fig.~1f) \; . 
$$
            
\begin{figure}[ht]
\centerline{
\hbox{
\includegraphics[width=8cm]{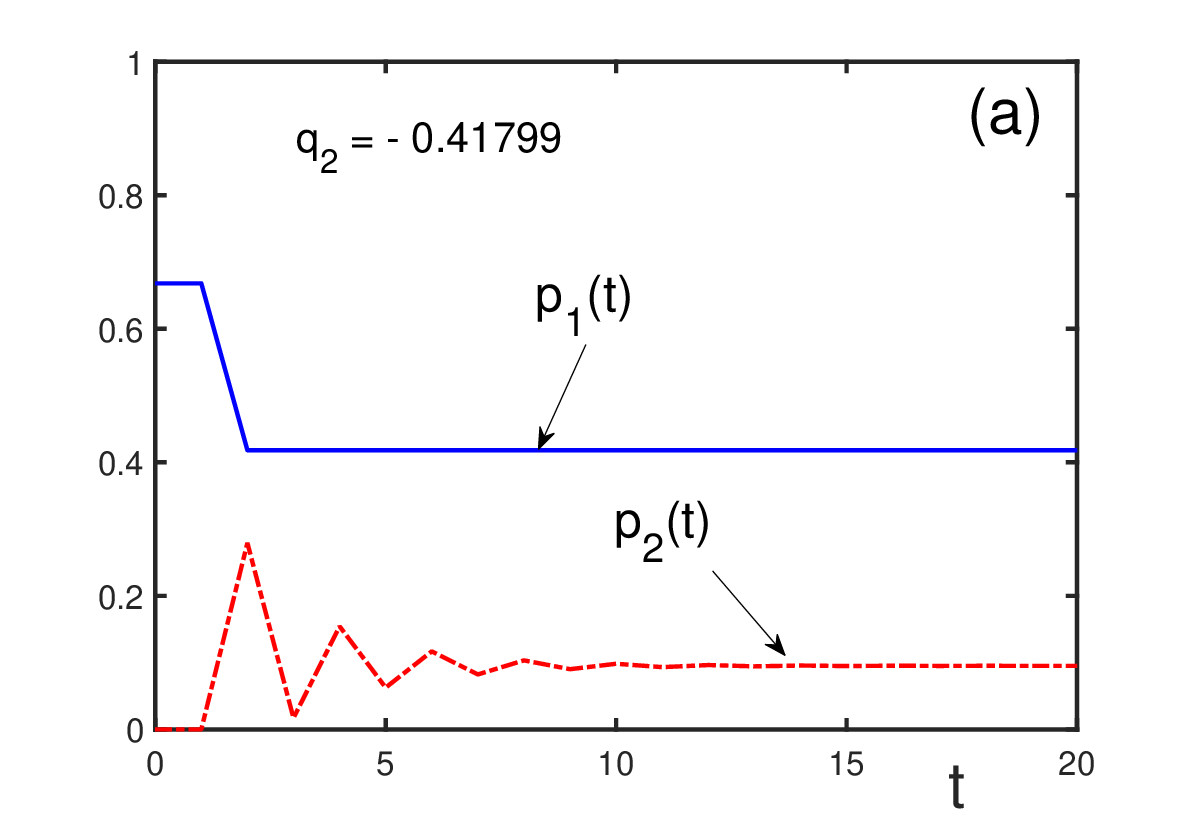} \hspace{1cm}
\includegraphics[width=8cm]{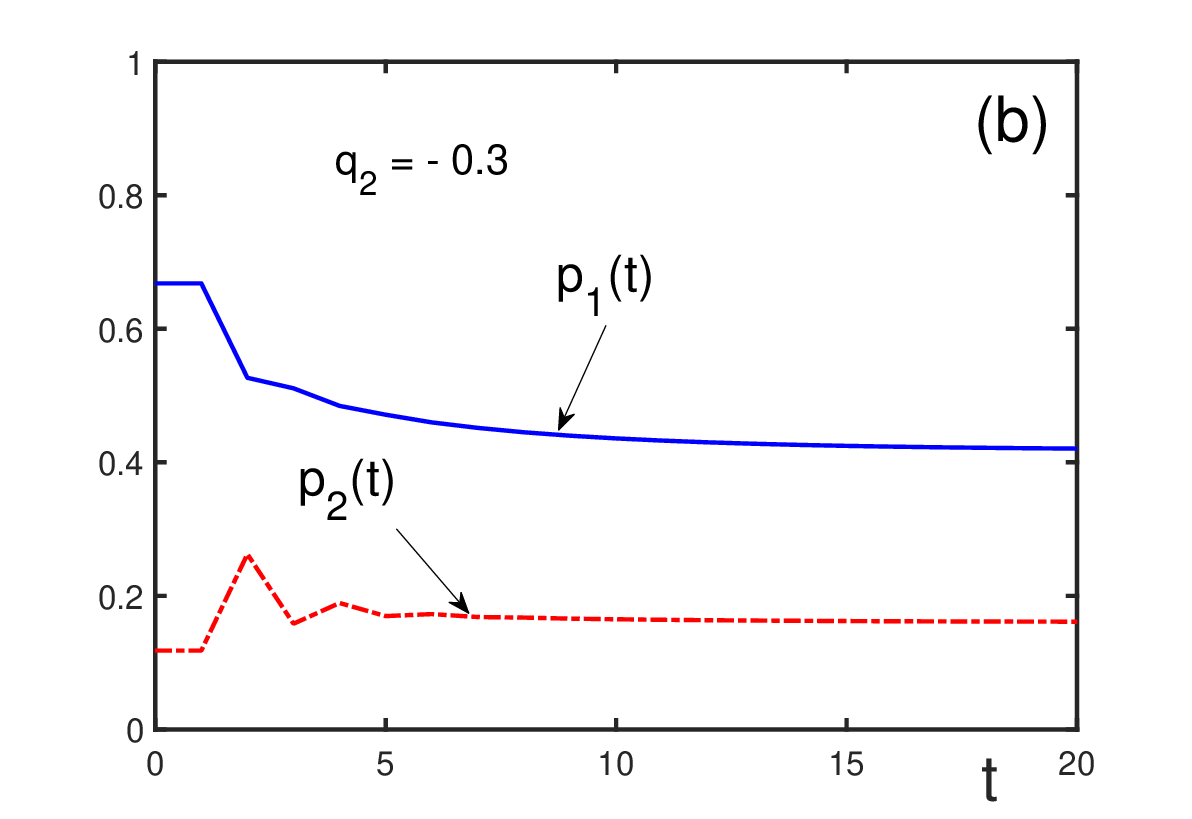} } }
\vskip 1cm
\centerline{
\hbox{
\includegraphics[width=8cm]{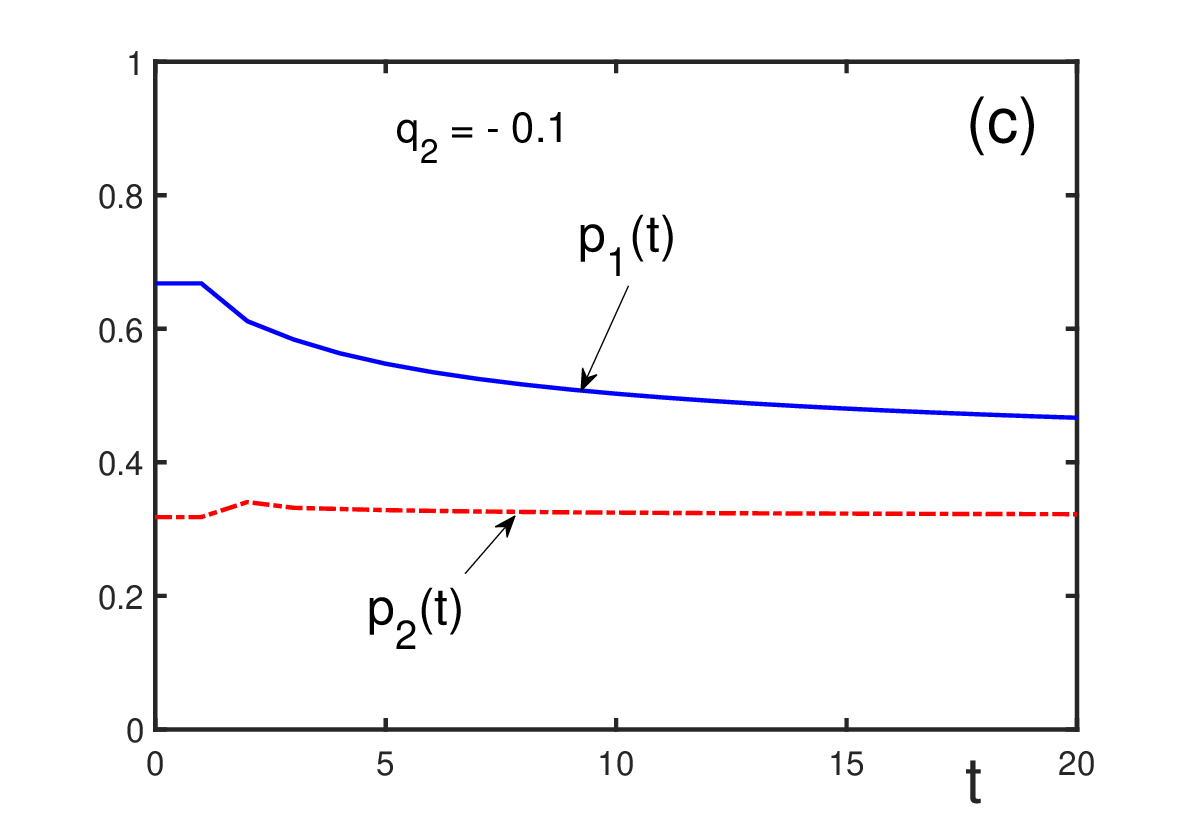} \hspace{1cm}
\includegraphics[width=8cm]{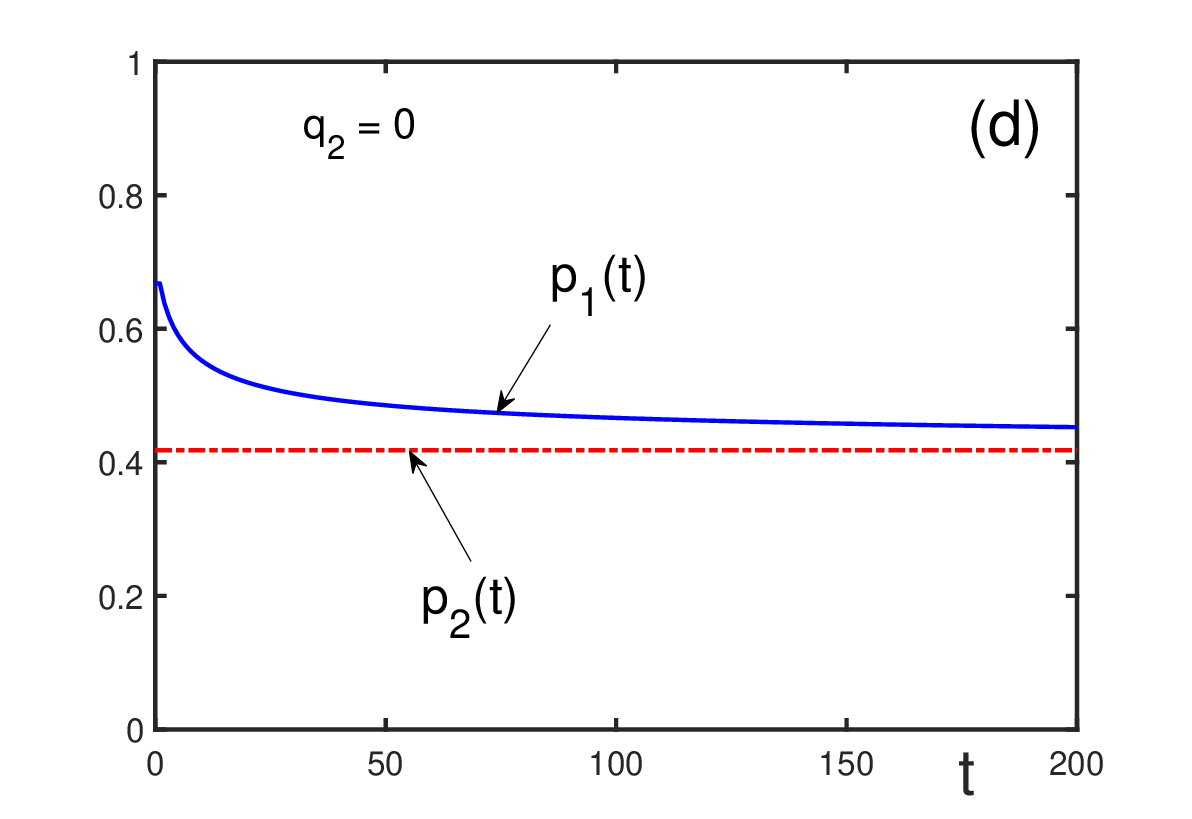} } }
\vskip 1cm
\centerline{
\hbox{
\includegraphics[width=8cm]{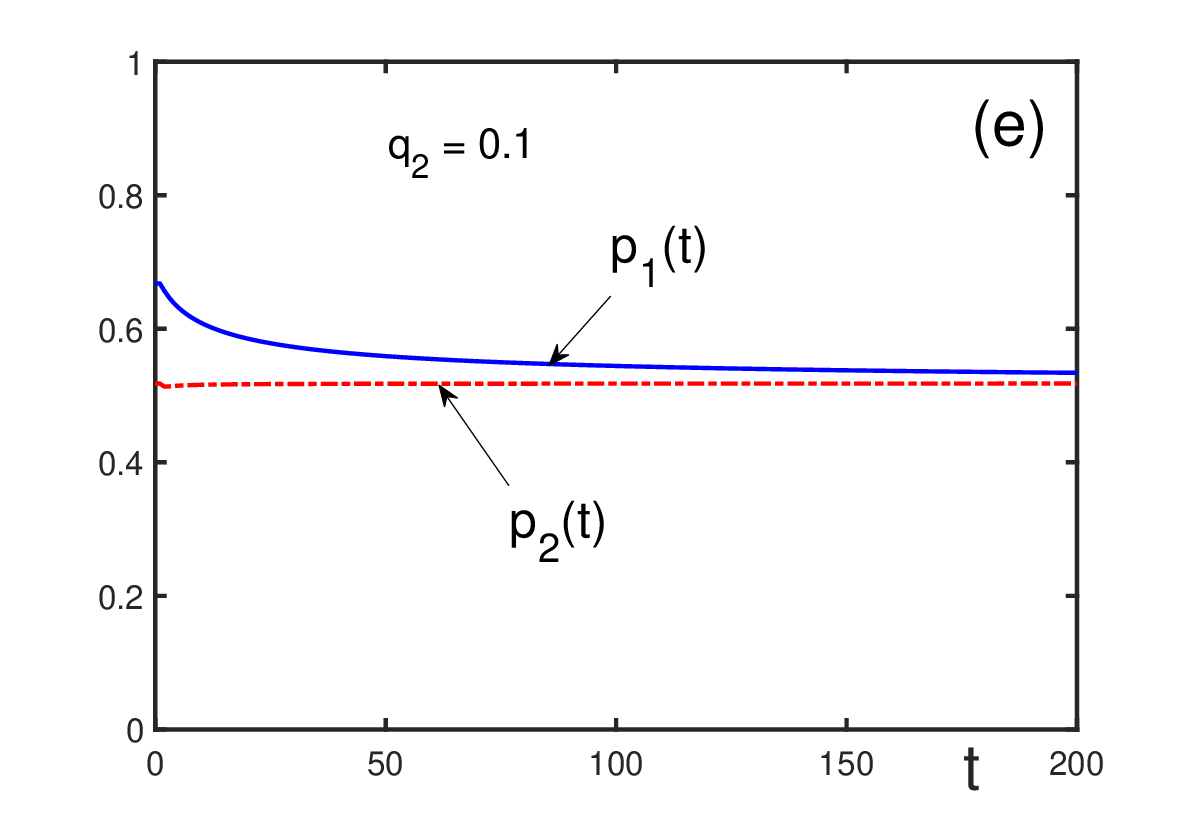} \hspace{1cm}
\includegraphics[width=8cm]{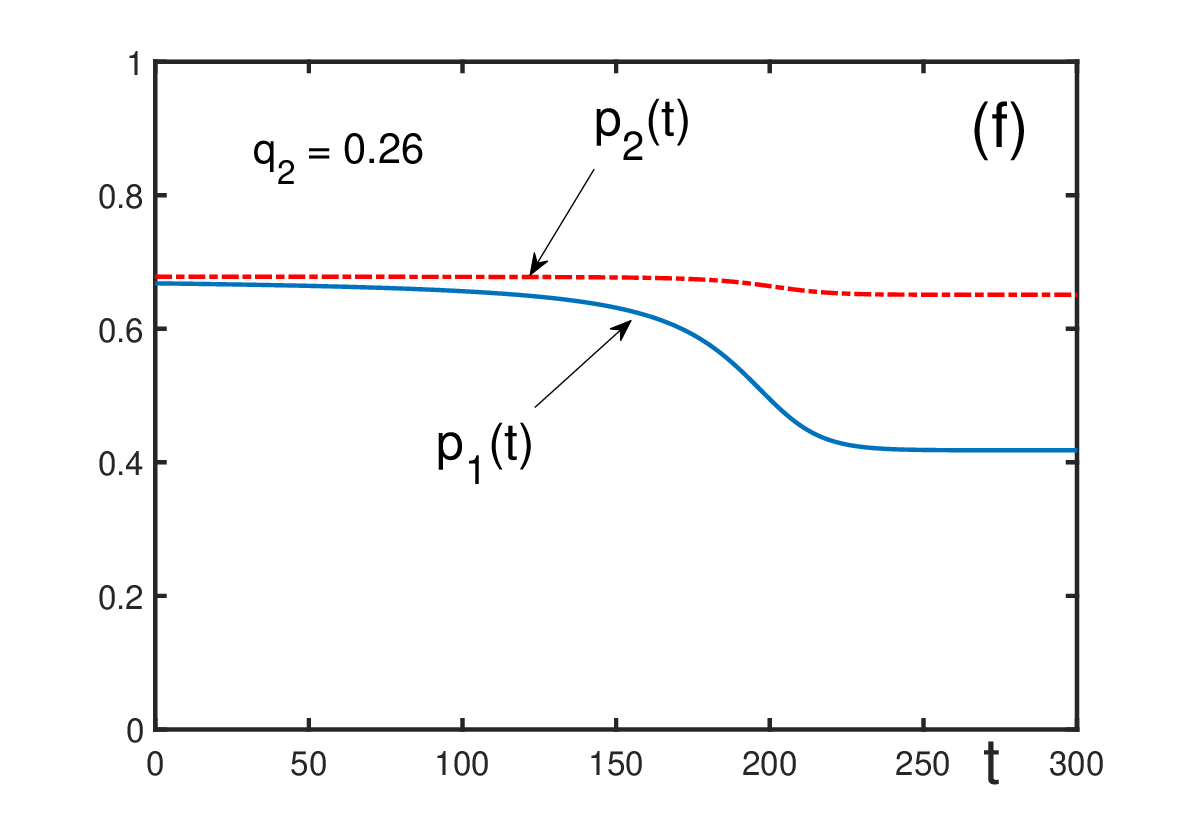} } }
\caption{\small
Probabilities $p_1(t)$ (solid line) and $p_2(t)$ (dashed-dotted line), describing the 
dynamics of affective decisions, in the absence of imitation, $\ep_1=\ep_2=0$, for 
the initial parameters $f=0.418$, $q_1=0.25$ and different $q_2$:
(a) $q_2=-0.41799$. Probability $p_1(t)\ra p_1^*=f=0.418$ and $p_2(t)\ra p_2^*=0.0952$;
(b) $q_2=-0.3$. Probability $p_1(t)\ra p_1^*=f=0.418$ and $p_2(t)\ra p_2^*=0.1607$;
(c) $q_2=-0.1$. Probability $p_1(t)\ra p_1^*=f=0.418$ and $p_2(t)\ra p_2^*=0.320$;
(d) $q_2=0$. Then $p_2(t)\equiv p_2(0)=f$ for $t\geq 0$, and $p_1(t)\ra p_1^*=f$
for $t\ra \infty$.
(e) $q_2=0.1$. Probabilities $p_1(t)\ra p_1^*$, $p_2(t)\ra p_2^*$, where $p_1^*=p_2^*=0.518$;
(f) $q_2=0.26>q_1=0.25$. Probabilities $p_1(t)\ra p_1^*=f=0.418$, $p_2(t)\ra p_2^*=0.651$. 
Note that for $q_2=q_1=0.25$, solutions $p_1(t)\equiv p_2(t)=0.668$ for $t\geq 0$.
}
\label{fig:Fig.1}
\end{figure}

\begin{figure}[ht]
\centerline{
\hbox{
\includegraphics[width=8cm]{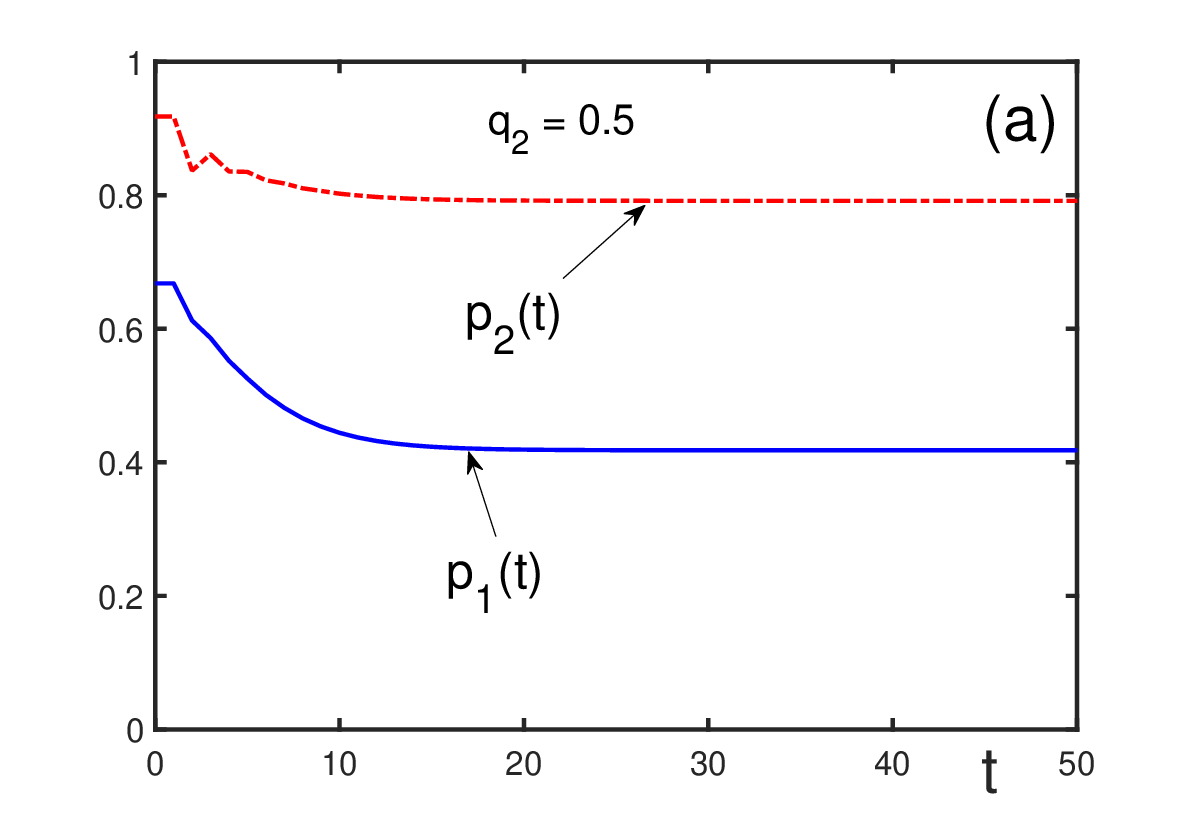} \hspace{1cm}
\includegraphics[width=8cm]{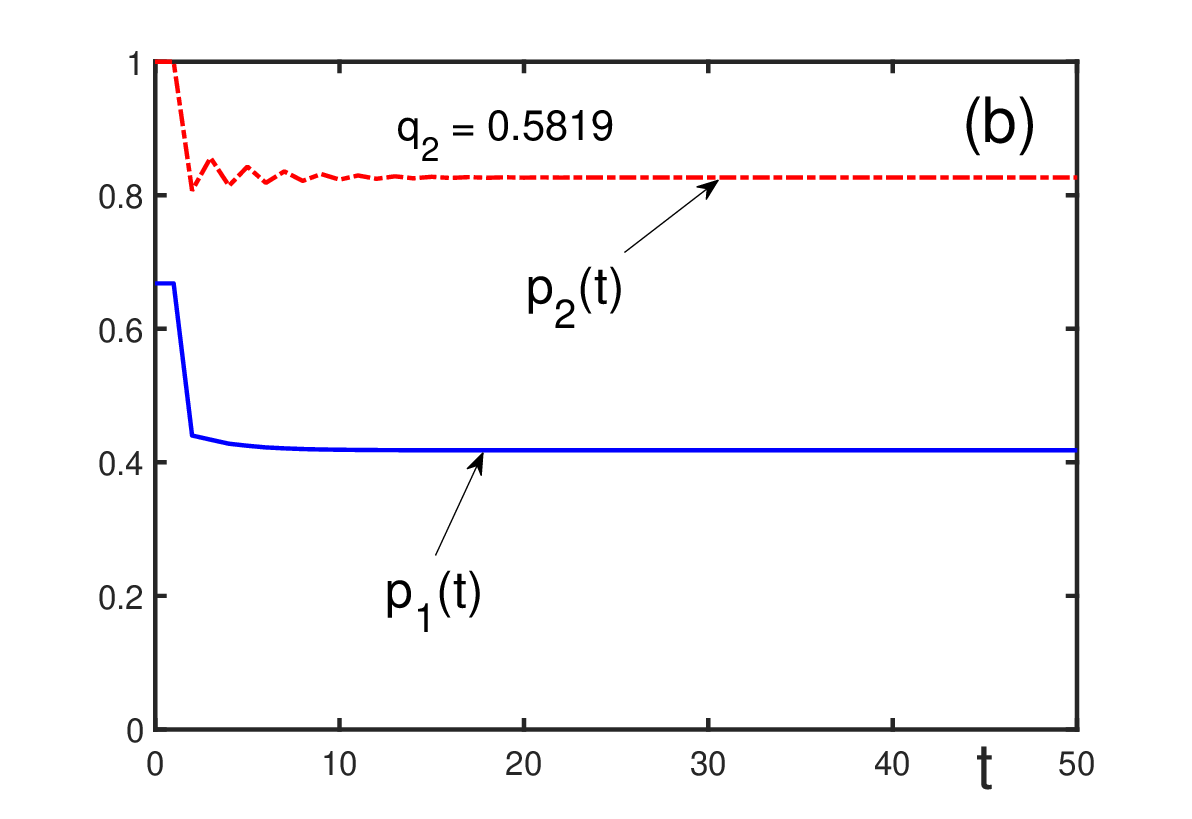} } }
\caption{\small
Probabilities $p_1(t)$ (solid line) and $p_2(t)$ (dashed-dotted line), as functions of time,
in the absence of imitation, $\ep_1=\ep_2=0$, for the initial parameters $f=0.418$, 
$q_1=0.25$, and different $q_2$:
(a) $q_2=0.5$. With increasing time, probabilities $p_1(t)\ra p_1^*=f=0.418$ and 
$p_2(t)\ra p_2^*=0.7916$.
(b) $q_2=0.5819$. With increasing time, probabilities $p_1(t)\ra p_1^*=f=0.418$ and 
$p_2(t)\ra p_2^*=0.8266$.
}
\label{fig:Fig.2}
\end{figure}

When the imitation effect is strong, so that $\varepsilon$ is close to one, the dynamics, 
depending on the initial attraction factor $q_2$, is similar to the case of the weak imitation 
effect, with the interchange of $p_1(t)$ and $p_2(t)$. For negative emotions,
\be
\label{79}
p_1(t) \; \ra \; p_1^* \; < \; f \; , \qquad p_2(t) \; \ra \;  f
\qquad
(-0.418<q_2<0 ) \;  ,
\ee
for neutral emotions,
\be
\label{80}
p_1(t) \; = \;  f \; , \qquad p_2(t) \; \ra \;  f
\qquad ( q_2 = 0 ) \;   ,
\ee
for weak positive emotions,
\be
\label{81}
p_1(t) \; \ra \; p^* \; > \; f \; , \qquad p_2(t) \; \ra \;  p^* \; > \; f
\qquad
( 0 < q_2 <0.25 ) \;   ,
\ee
and for strong positive emotions,
\be
\label{82}
p_1(t) \; \ra \; p_1^* \; > \; f \; , \qquad p_2(t) \; \ra \;  f
\qquad
( 0.25 < q_2 < 0.573 ) \;   .
\ee

These regimes, to some extent, are really similar to those in the case of weak imitation, 
summarized in Eqs. (\ref{75}) to (\ref{78}), with the interchange between $p_1(t)$ and $p_2(t)$.
However, the similarity is not absolutely complete, since the memories $M_1(t)$ and $M_2(t)$ 
have different forms, and are not symmetric with respect to the interchange between $p_1(t)$ 
and $p_2(t)$. Therefore, as compared to the case of absent imitation, under strong herding 
there exists one more regime in the range $0.573 < q_2 < o.582$, where $p_1(t)$ permanently 
oscillates, while $p_2(t)$ tends to $f$. The arising oscillations in $p_1(t)$ are caused by 
strong propensity to imitation, when the agents with long-term memory are strongly influenced 
by the behavior of agents with short-term memory. Permanent oscillations imply the difficulty 
of decision makers for coming to a firm decision because of persistent deliberations. The 
existing regimes are shown in Figs. $3$ and $4$.

\begin{figure}[ht]
\centerline{
\hbox{
\includegraphics[width=8cm]{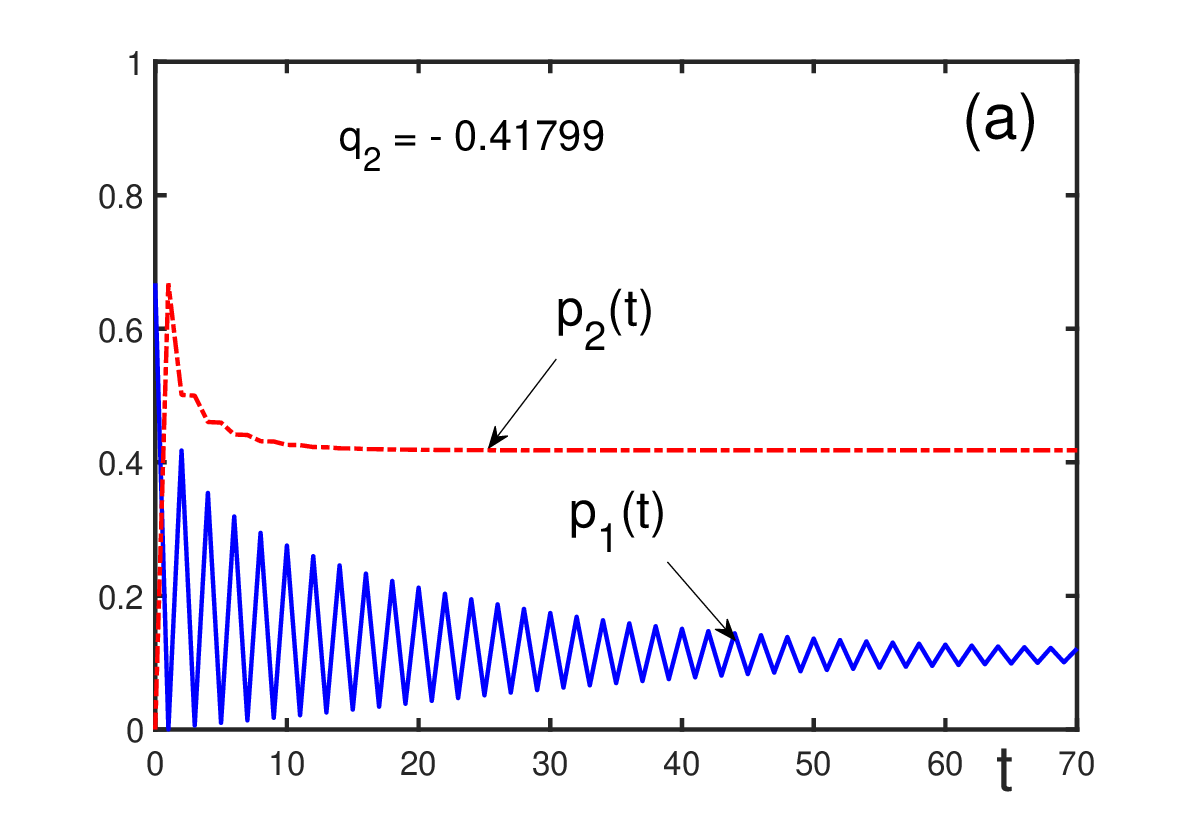} \hspace{1cm}
\includegraphics[width=8cm]{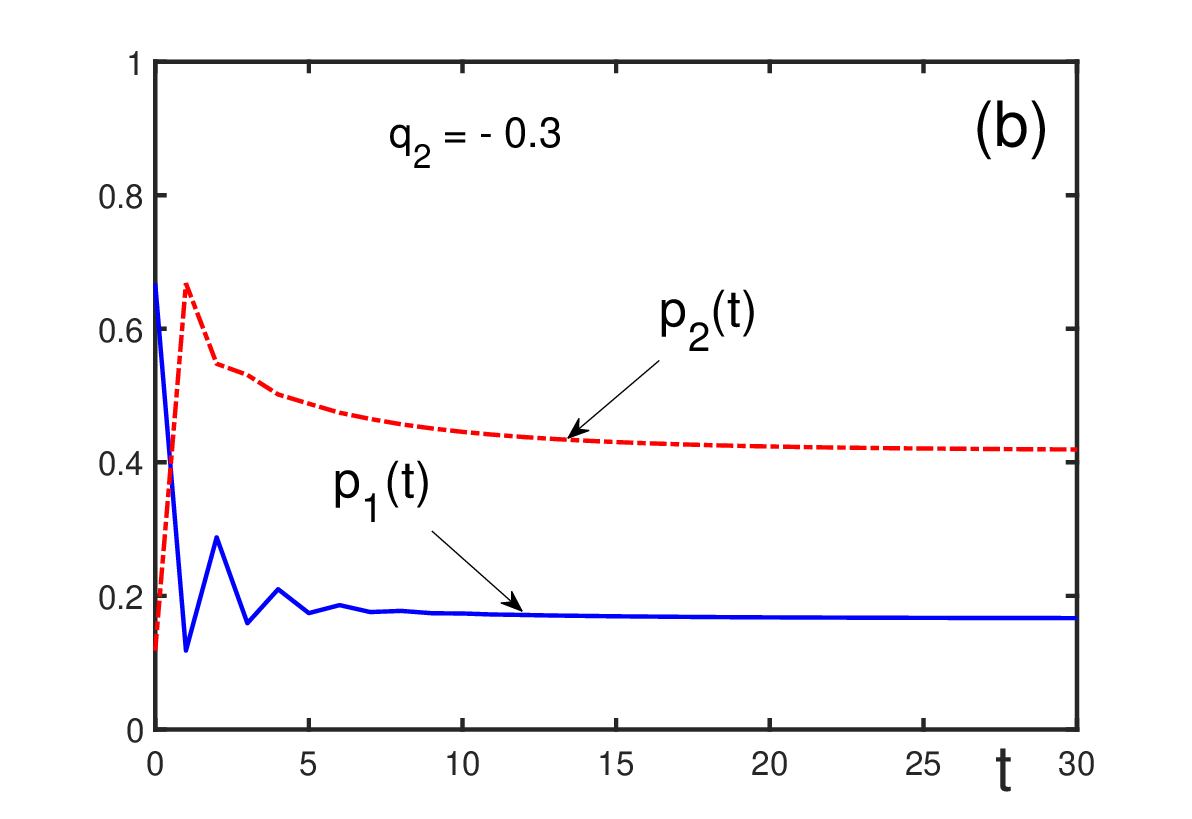} } }
\vskip 1cm
\centerline{
\hbox{
\includegraphics[width=8cm]{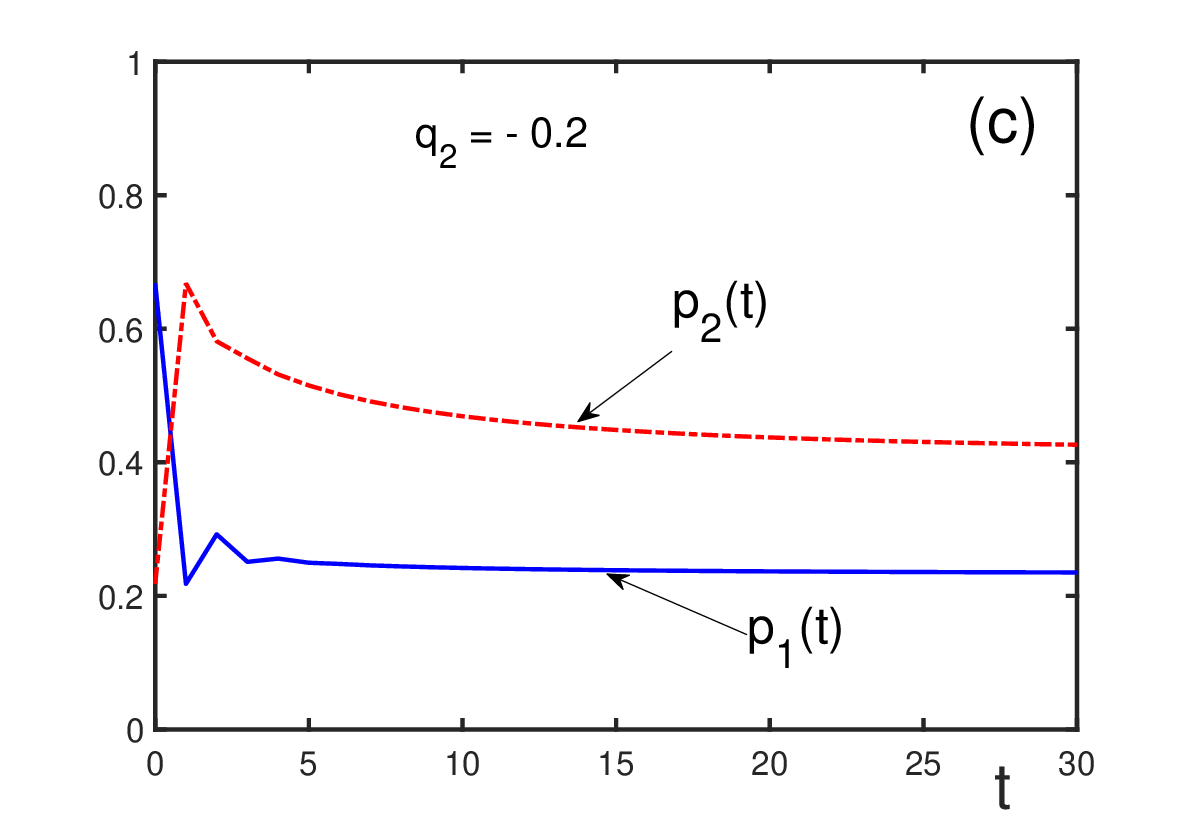} \hspace{1cm}
\includegraphics[width=8cm]{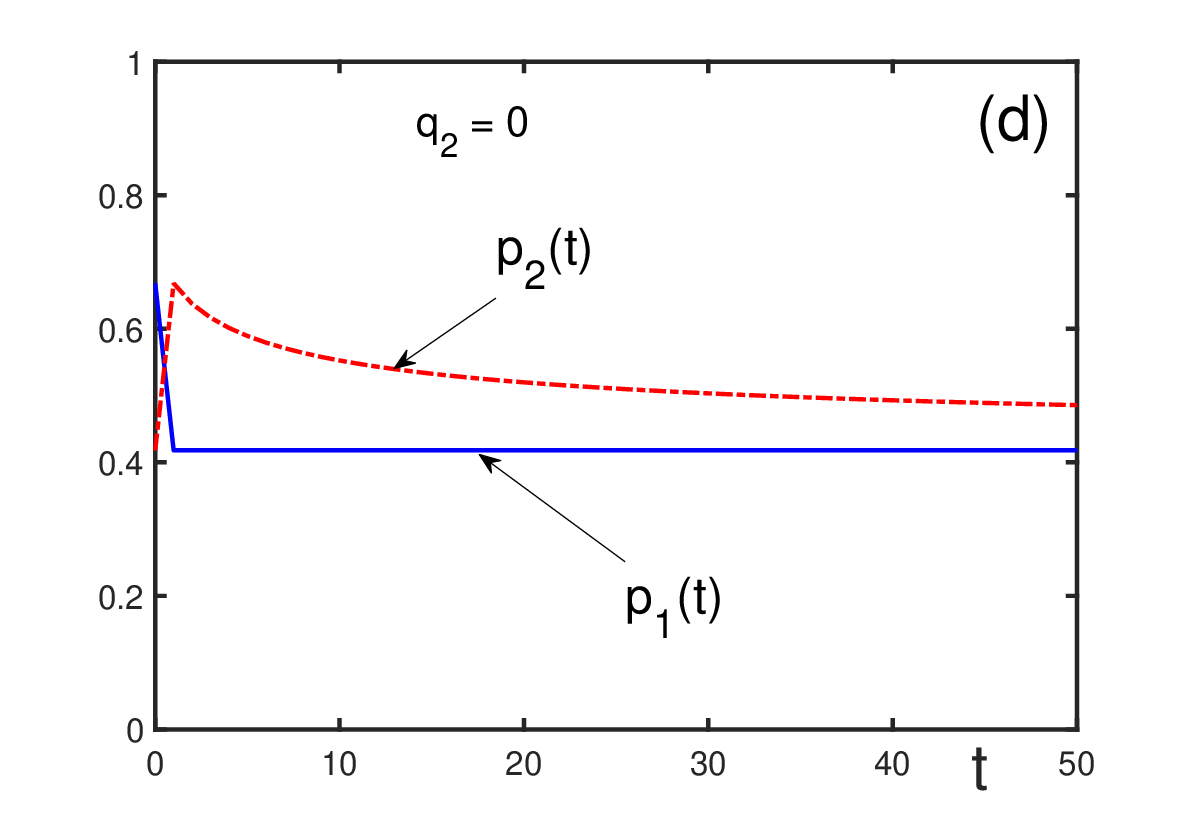} } }
\vskip 1cm
\centerline{
\hbox{
\includegraphics[width=8cm]{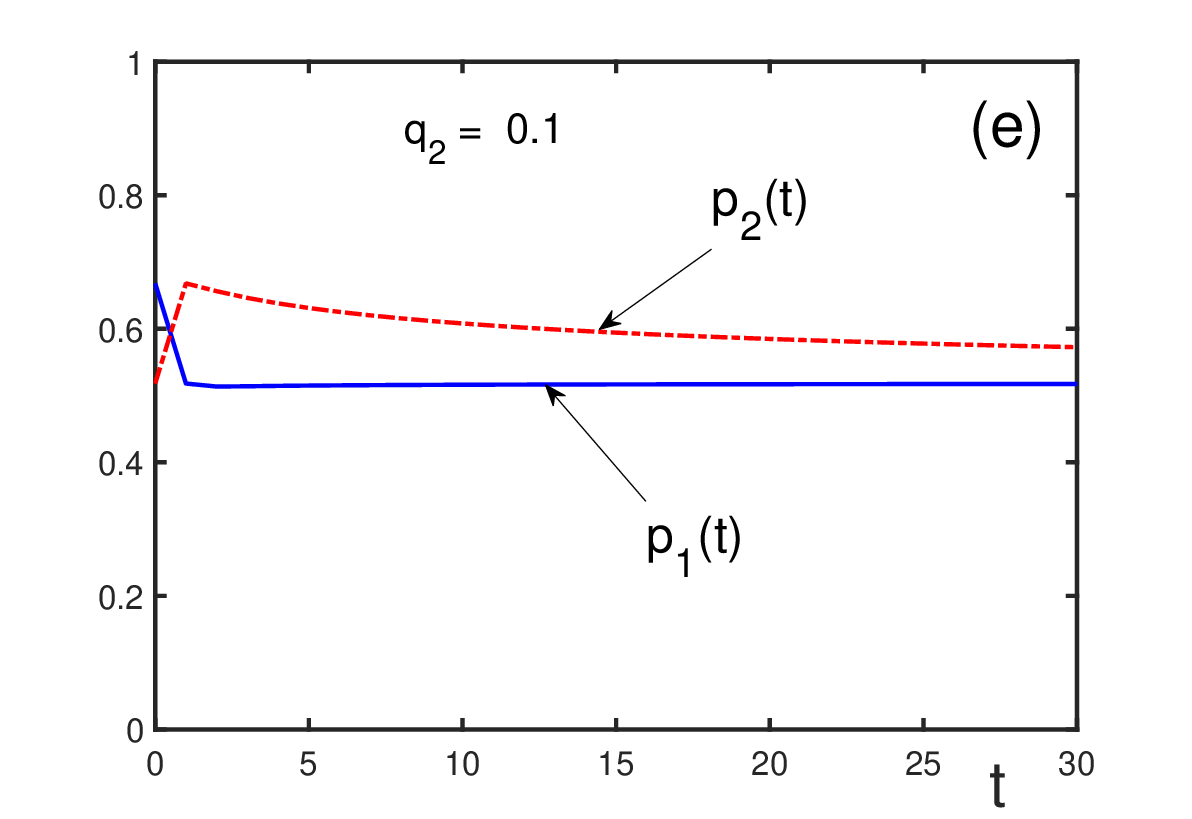} \hspace{1cm}
\includegraphics[width=8cm]{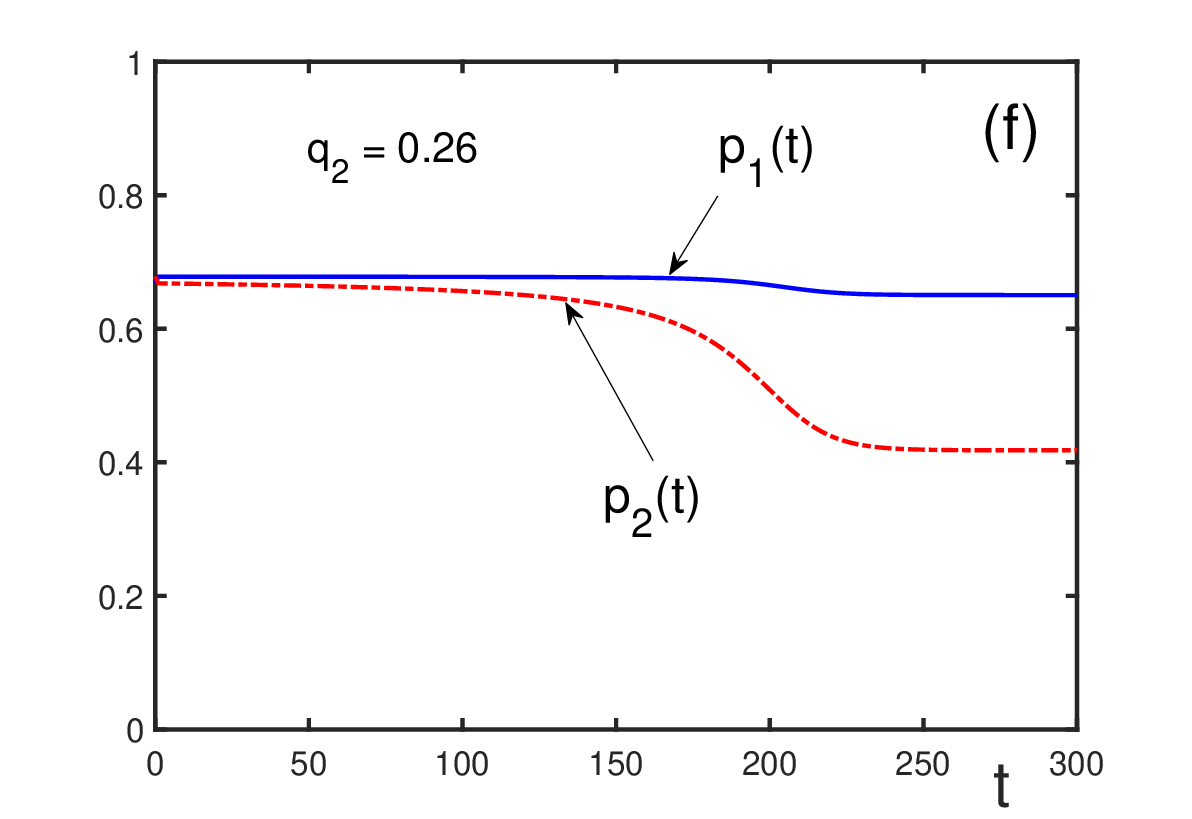} } }
\caption{\small
Dynamics of the probabilities $p_1(t)$ (solid line) and $p_2(t)$ (dashed-dotted line)
under strong imitation effect, $\ep_1=\ep_2=1$, for the initial parameters $f=0.418$, $q_1=0.25$ 
and different $q_2$:
(a) $q_2=-0.41799$. Probabilities $p_2(t)\ra p_2^*=f=0.418$ and $p_1(t)\ra p_1^*=0.1109$;
(b) $q_2=-0.3$. Probabilities $p_2(t)\ra p_2^*=f=0.418$ and $p_1(t)\ra p_1^*=0.1664$;
(c) $q_2=-0.2$. Probabilities $p_2(t)\ra p_2^*=f=0.418$ and $p_1(t)\ra p_1^*=0.2339$;
(d) $q_2=0$. Probabilities $p_2(t)\ra p_2^*=f$ and $p_1(t)\equiv p_1(1)=f$;
(e) $q_2=0.1$. Probabilities $p_2(t)\ra p_2^*=0.518$ and $p_1(t)\ra p_1^*=0.518$; 
(f) $q_2=0.26>q_1=0.25$. Probabilities $p_2(t)\ra p_2^*=f=0.418$ and $p_1(t)\ra p_1^*=0.6505$. 
Note that for $q_2=q_1=0.25$, the probabilities $p_1(t)\equiv p_2(t)=0.668$ for $t\geq 0$.
}
\label{fig:Fig.3}
\end{figure}

\begin{figure}[ht]
\centerline{
\hbox{
\includegraphics[width=8cm]{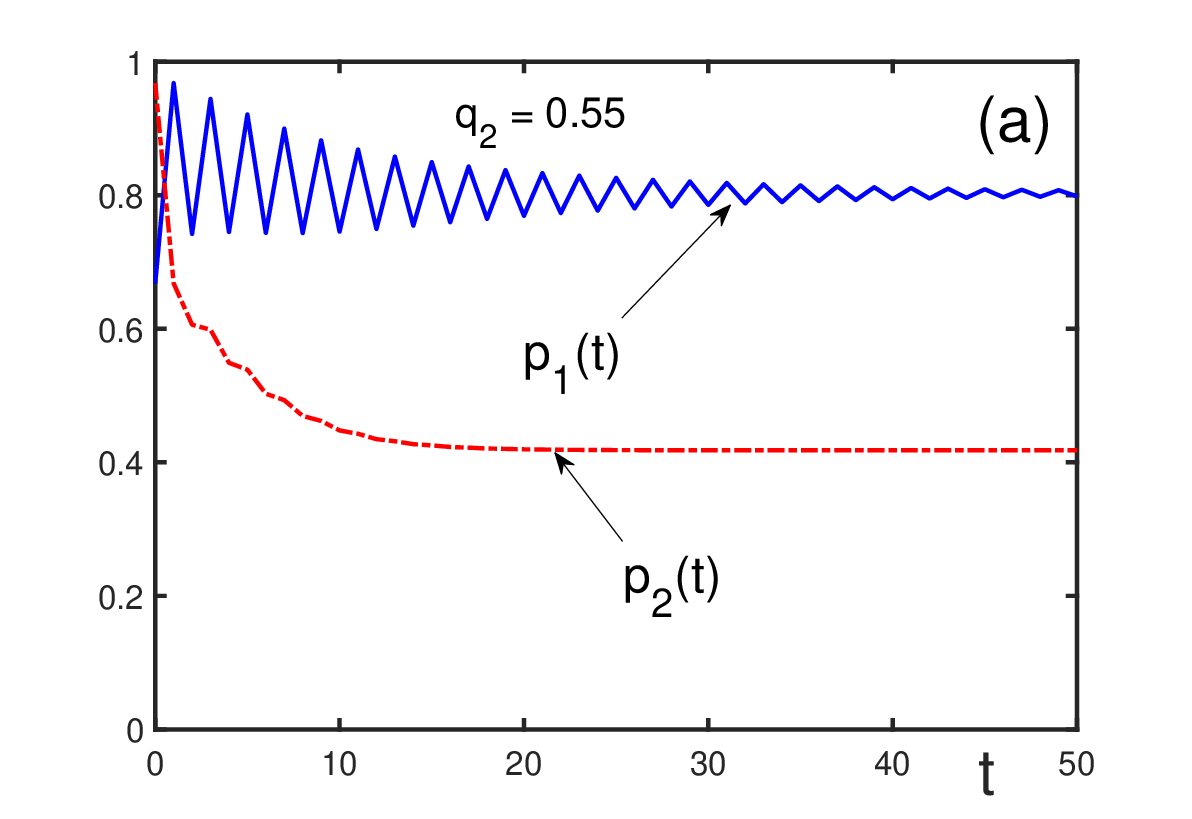} \hspace{1cm}
\includegraphics[width=8cm]{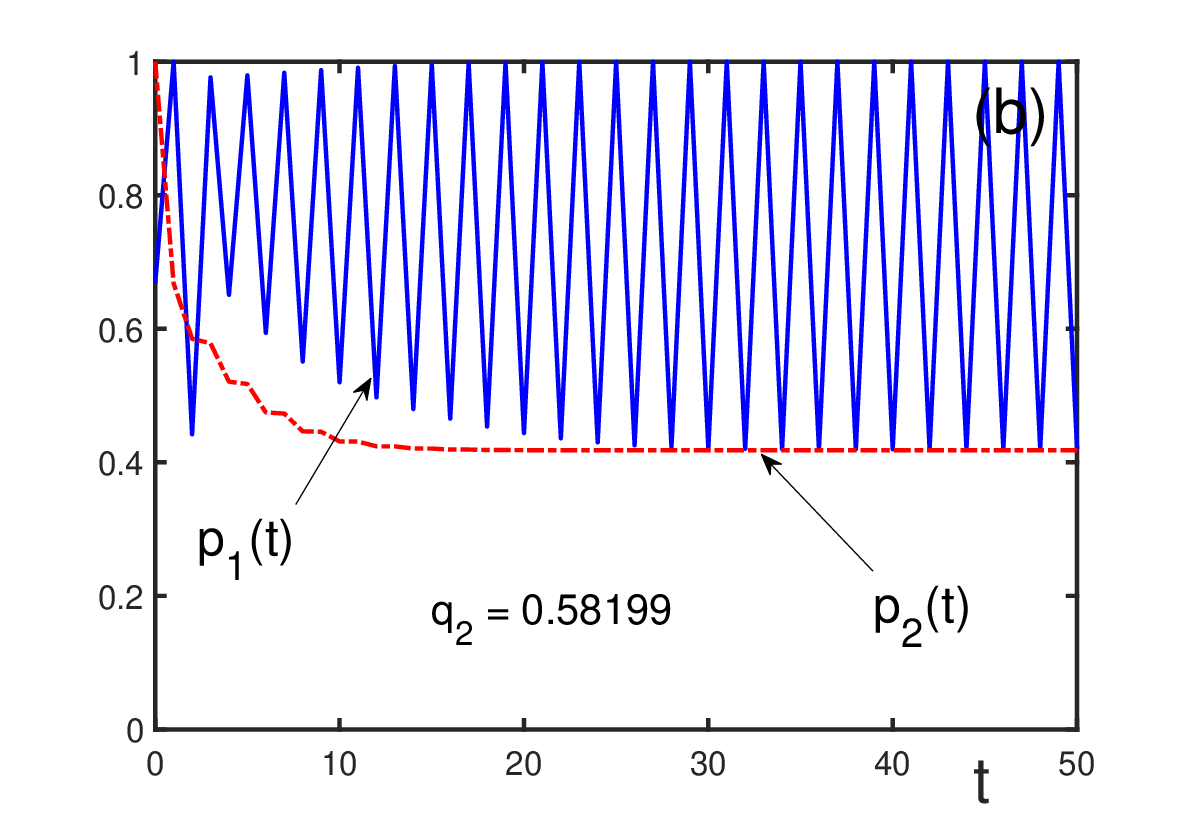} } }
\caption{\small
Probabilities $p_1(t)$ (solid line) and $p_2(t)$ (dashed-dotted line) under strong imitation 
effect, $\ep_1=\ep_2=1$, for the initial parameters $f=0.418$, $q_1=0.25$, and different $q_2$:
(a) $q_2=0.55$. Solutions $p_2(t)\ra p_2^*=f$ and $p_1(t)\ra p_1^*=0.8028$; 
(b) $q_2=0.58199$. When $t\ra\infty$, then $p_2(t)\ra p_2^*=f$ and $p_1(t)$ permanently 
oscillates.}
\label{fig:Fig.4}
\end{figure}

The most important point that needs to be emphasized is that the exchange of information
results in the diminishing influence of emotions. For both types of decision makers, with 
long as well as with short-term memory, the final attraction factor is smaller by magnitude
than the initial attraction factor, $|q_i^*| \leq |q_i|$. For the agents with long-term 
memory, in the majority of cases the attraction factor becomes zero, which signifies the 
disappearance of emotional influence. However, the final probabilities $p_1^*$ and $p_2^*$ 
for different groups are usually different, meaning that there is no consensus between the 
different groups forming the society. 

The attenuation of the modulus of the attraction factor, hence of the emotional influence,
with the repeated decision making by a group of agents exchanging information, is a well
established empirical fact confirmed in a number of experiments \cite{Kuhberger_2001,
Charness_2002,Blinder_2005,Cooper_2005,Sutter_2005,Tsiporkova_2006,Charness_2007,
Charness_Rigotti_2007,Chen_2009,Liu_2009,Charness_2010}.

The diminishing role of emotions means that the absolute value of the attraction factor 
$\;|q_j(A_n,t)\;|$ diminishes, hence the deviation of the probability $p_i(A_n,t)$ from the 
utility factor $f_j(A_n)$ becomes smaller. But when the probability for one alternative, 
e.g. $A_1$, diminishes, the probability for another alternative $A_2$ rises due to the 
normalization conditions (50) of Sec. 3.2. To make this transparent, let us consider an 
example, where it is required to choose between two alternatives, one is war (alternative $A_1$), 
the other is peace (alternative $A_2$). Suppose, due to propaganda or other reasons, the
population is overwhelmingly in favor of war, so that $p_j(A_1)$ rises fast. This is what 
one names military polarization. At the same time, the probability for peace $p_j(A_2)$ 
falls down, and vice versa. Thus, talking on polarization of opinions, one has to specify
which alternatives are kept in mind. In the general sense, polarization can be understood
as the situation when the probability of one alternative is large, while the probability 
of another alternative is small \cite{Galam_2012}. Such an essential variation of probabilities 
can be induced by the variation of the utility factor or attraction factor. For instance, 
in Figs. 7 to 9, the case is shown when $p_j(A_1,t) = p_j(t)$ grows to one, hence 
$p_j(A_2,t) =1 - p_j(t)$ falls down to zero.

\section{Regulated decision making}

The natural question that arises is whether it would be possible to regulate the decisions of 
the society agents so that all members would come to the required prescribed conclusion? This 
could be possible by resorting to machine learning techniques \cite{Alpaydin_72}, by regulating 
the utility of the given alternatives. For instance, the utility factor for a group $i$ can be 
varied according to the law
\be  
\label{83}
 f_i(t) \; = \; f_i\; \exp\{ K_i(t) \} \;  ,
\ee
or, more correctly, by keeping in mind the normalization of the utility factors, by the law
\begin{eqnarray}
\label{84}
f_i^R(t) \; = \; \left\{ \begin{array}{ll}
f_i(t) \; , ~ & ~ f_i(t) \; \leq \; 1 \\
1 \; , ~ & ~ f_i(t) \; > \; 1 \; .
\end{array}
\right.
\end{eqnarray}
The utility can be regulated by a gradient-descent equation resulting in
\be
\label{85}
K_i(t+1) \; = \; K_i(t) - \al_i \; [ \; p_i(t) - p_i^* \; ] \;   ,
\ee
with the initial condition 
\be
\label{86}
 K_i(0) \; = \; 0 \;  ,
\ee
where $\alpha_i$ is a learning rate and $p_i^*$ is the decision  probability to which the 
agents of the group $i$ are forced to converge. The learning rate is chosen so that to 
guarantee the procedure stability \cite{Alpaydin_72}. For the case of the binary choice, 
equation (\ref{57}) becomes
$$
p_1(t+1) \; = \; ( 1 - \ep_1) \; R_{01} \left[ \; f_1^R(t) + q_1(t) \; \right] 
+ \ep_1\; R_{01} \left[ \; f_2^R(t) + q_2(t) \; \right] \;   ,
$$
\be
\label{87}
p_2(t+1) \; = \; ( 1 - \ep_2) \; R_{01} \left[ \; f_2^R(t) + q_2(t) \; \right] 
+ \ep_2\; R_{01} \left[ \; f_1^R(t) + q_1(t) \; \right] \;   ,
\ee
where the retract function is
\begin{eqnarray}
\label{88}
R_{01}[\; z\; ] \; = \; \left\{ \begin{array}{ll}
0 \; , ~ & ~ z \; < \; 0 \\
z \; , ~ & ~ 0 \leq z \leq 1 \\
1 \; , ~ & ~ z \; > \; 1 \; .
\end{array} \right.
\end{eqnarray}

As examples, we show in Figs. 5 to 6 how the agents can be forced to converge to the 
consensual limit $p_i^* = 0.418$ equal to the utility factor $f_i = 0.418$ in the case 
of the Allais paradox. The initial conditions are different due to the presence of emotions
that are different for different agents, but the final decisions are the same for all agents, 
as is prescribed by the gradient-descent equations (\ref{85}). In Figs. 5 to 6, the imitation 
effect is neglected. As we have checked, small $\varepsilon_i$ does not change much the 
resulting behavior. Small learning rate forces the decisions to slowly converge to the 
consensual limit. Too large learning rate makes the process unstable, as is should be 
\cite{Alpaydin_72}, so that the value of the learning rate is prescribed by the 
conditions of the process stability.  

In order to show that the final decision probability can be chosen arbitrarily, in Figs. 
7 to 9 the prescribed final probability is taken as $p_i^* = 1$, which implies that all agents
are forced to choose the first alternative with the probability one, while without regulations
their choices are different.

\section{Discussion}

A novel type of networks is developed, dynamic probabilistic decision network. The nodes of
the network are intelligent agents making probabilistic decisions based on the rational 
evaluation of the given alternatives as well as taking into account the agent's emotions 
associated with these alternatives. At the first step, the agents make decisions independently 
from each other. After the initial stage of individual decision making, they start 
communicating between themselves by exchanging information on their choices and by imitating 
the behavior of others. The network in general, is non-homogeneous, comprising different 
agents, possessing different types of memory, some having long-term memory, while others, 
short-term memory. Such probabilistic dynamic networks can represent different realistic 
social systems, either human and other biological societies, or neural networks. 

The graphical representation of temporal networks is essentially more complicated than that 
of static networks composed of static graphs, since temporal networks, generally, require
to characterize their topological structure at each moment of time as well as the network 
dynamics with varying time. In the case when the interactions between the network agents
is of long-range, the topological structure is not important and it is possible to represent
the network operation by a set of temporal slices, where layers correspond to different 
moments of time \cite{Holme_73}. As an example, the class of networks, considered in the 
present paper, is represented in Fig. 10 for the case of three groups of agents and three 
temporal slices.    

The probabilistic network operation is illustrated by considering the Allais paradox in both
the static and dynamic pictures. First, it is shown that in the affective decision making the 
Allais paradox finds a simple and natural resolution. Then the consideration is generalized 
to multi-step dynamics in the frame of probabilistic dynamic system, whose agents exchange 
information with each other and exhibit the imitation effect. When the propensity to imitation 
is strong, then because of the presence of agents with short-term memory, there can appear 
an oscillatory dynamics implying permanent hesitations of decision makers. However, when the 
propensity to imitation is moderate or weak, the role of emotions in collective decision 
making diminishes with time, so that the agents are more inclined to make their choice on 
rational grounds. 

By employing machine learning techniques, it is possible to regulate the value of the final 
probability thus forcing the agents to chose the alternatives as required.    

The developed theory of probabilistic affective decision network can find application in 
the description of complex social networks, in characterizing neuronal brain networks, and 
in the creation of artificial intelligence operating similarly to emotional biological beings. 
     
Applications of the model for other possible particular cases and the analysis of its 
predictions for some concrete situations will be the topic of future research. The aim
of the present paper has been to introduce a new model that takes into consideration realistic 
features existing in any real-life society, where decisions are made on the basis of several
attributes: evaluation of utility, influence of emotions, imitation effect, exchange of 
information, existence of memory, dynamics of repeated decisions, and probabilistic nature
of decision process. These are compulsory features of any intelligent complex society, which
are necessary to take into account for a realistic description. At the present time, there are 
no other models taking account of all these features. Complex systems require complex models.

\vskip 2mm

\section*{Declaration of competing interests}

We have nothing to declare.

\section*{Funding}
 
This research did not receive any specific grant from funding agencies in the public, 
commercial, or not-for-profit sectors.

\section*{Competing Interests}

The authors have no conflicts to disclose. 

\section*{Authors' contribution}

The study conception and design were done by V.I.Y. and E.P.Y. Material preparation, data 
collection, and analysis were performed by V.I.Y. and E.P.Y. The first draft of the manuscript 
was written by V.I.Y. and all authors commented on previous versions of the manuscript. 
Numerical calculations were accomplished by E.P.Y. All authors read and approved the final 
manuscript.

\newpage

\begin{center}
{\Large{\bf Figure Captions}}
\end{center}

\vskip 1cm
{\bf Fig. 1}.
Probabilities $p_1(t)$ (solid line) and $p_2(t)$ (dashed-dotted line), describing the 
dynamics of affective decisions, in the absence of imitation, $\ep_1=\ep_2=0$, for 
the initial parameters $f=0.418$, $q_1=0.25$ and different $q_2$:
(a) $q_2=-0.41799$. Probability $p_1(t)\ra p_1^*=f=0.418$ and $p_2(t)\ra p_2^*=0.0952$;
(b) $q_2=-0.3$. Probability $p_1(t)\ra p_1^*=f=0.418$ and $p_2(t)\ra p_2^*=0.1607$;
(c) $q_2=-0.1$. Probability $p_1(t)\ra p_1^*=f=0.418$ and $p_2(t)\ra p_2^*=0.320$;
(d) $q_2=0$. Then $p_2(t)\equiv p_2(0)=f$ for $t\geq 0$, and $p_1(t)\ra p_1^*=f$
for $t\ra \infty$.
(e) $q_2=0.1$. Probabilities $p_1(t)\ra p_1^*$, $p_2(t)\ra p_2^*$, where $p_1^*=p_2^*=0.518$;
(f) $q_2=0.26>q_1=0.25$. Probabilities $p_1(t)\ra p_1^*=f=0.418$, $p_2(t)\ra p_2^*=0.651$. 
Note that for $q_2=q_1=0.25$, solutions $p_1(t)\equiv p_2(t)=0.668$ for $t\geq 0$.

\vskip 1cm
{\bf Fig. 2}.
Probabilities $p_1(t)$ (solid line) and $p_2(t)$ (dashed-dotted line), as functions of time,
in the absence of imitation, $\ep_1=\ep_2=0$, for the initial parameters $f=0.418$, 
$q_1=0.25$, and different $q_2$:
(a) $q_2=0.5$. With increasing time, probabilities $p_1(t)\ra p_1^*=f=0.418$ and 
$p_2(t)\ra p_2^*=0.7916$.
(b) $q_2=0.5819$. With increasing time, probabilities $p_1(t)\ra p_1^*=f=0.418$ and 
$p_2(t)\ra p_2^*=0.8266$.

\vskip 1cm
{\bf Fig. 3}.
Dynamics of the probabilities $p_1(t)$ (solid line) and $p_2(t)$ (dashed-dotted line)
under strong imitation effect, $\ep_1=\ep_2=1$, for the initial parameters $f=0.418$, $q_1=0.25$ 
and different $q_2$:
(a) $q_2=-0.41799$. Probabilities $p_2(t)\ra p_2^*=f=0.418$ and $p_1(t)\ra p_1^*=0.1109$;
(b) $q_2=-0.3$. Probabilities $p_2(t)\ra p_2^*=f=0.418$ and $p_1(t)\ra p_1^*=0.1664$;
(c) $q_2=-0.2$. Probabilities $p_2(t)\ra p_2^*=f=0.418$ and $p_1(t)\ra p_1^*=0.2339$;
(d) $q_2=0$. Probabilities $p_2(t)\ra p_2^*=f$ and $p_1(t)\equiv p_1(1)=f$;
(e) $q_2=0.1$. Probabilities $p_2(t)\ra p_2^*=0.518$ and $p_1(t)\ra p_1^*=0.518$; 
(f) $q_2=0.26>q_1=0.25$. Probabilities $p_2(t)\ra p_2^*=f=0.418$ and $p_1(t)\ra p_1^*=0.6505$. 
Note that for $q_2=q_1=0.25$, the probabilities $p_1(t)\equiv p_2(t)=0.668$ for $t\geq 0$.

\vskip 1cm
{\bf Fig. 4}.
Probabilities $p_1(t)$ (solid line) and $p_2(t)$ (dashed-dotted line) under strong imitation 
effect, $\ep_1=\ep_2=1$, for the initial parameters $f=0.418$, $q_1=0.25$, and different $q_2$:
(a) $q_2=0.55$. Solutions $p_2(t)\ra p_2^*=f$ and $p_1(t)\ra p_1^*=0.8028$; 
(b) $q_2=0.58199$. When $t\ra\infty$, then $p_2(t)\ra p_2^*=f$ and $p_1(t)$ permanently 
oscillates.

\vskip 1cm
{\bf Fig. 5}.
Probabilities $p_1(t)$ (solid line) and $p_2(t)$ (dashed line) for parameters 
$\ep_1=\ep_2=0$, $q_1=0.25$, $q_2=0.2$, $f_1=f_2=0.418$, $p^*=0,418$, and different $\al$:
(a) $\al=0$. For $t\ra\infty$, probabilities $p_1(t)\ra \overline p_1^*$ and 
$p_2(t)\ra \overline p_2^*$, where $\overline p_1^*=\overline p_2^*=0.618$.
(b) $\al=0.5$. Both functions, $p_1(t)\ra p^*$ and $p_2(t)\ra p^*$, tend to the same 
limit, $p^*=f_1=f_2=0.418$, when $t\ra\infty$.

\vskip 1cm
{\bf Fig. 6}.
Probabilities $p_1(t)$ (solid line) and $p_2(t)$ (dashed line) for parameters 
$\ep_1=\ep_2=0$, $q_1=0.25$, $q_2=0.4$, $f_1=f_2=0.418$, $p^*=0,418$, and different $\al$:
(a) $\al=0$. For $t\ra\infty$, probabilities $p_1(t)\ra \overline p_1^*=p^*=f_1=0.418$ and 
$p_2(t)\ra \overline p_2^*=0.7408$.
(b) $\al=0.5$. Both functions, $p_1(t)\ra p^*$ and $p_2(t)\ra p^*$, tend to the same 
limit, $p^*=f_1=0.418$, when $t\ra\infty$.

\vskip 1cm
{\bf Fig. 7}.
Probabilities $p_1(t)$ (solid line) and $p_2(t)$ (dashed line) for parameters 
$\ep_1=\ep_2=0$,  $q_1=0.25$, $q_2=0.1$, $f_1=f_2=0.418$, $p^*=1$, and different $\al$:
(a) $\al=0$. For $t\ra\infty$, probabilities $p_1(t)\ra \overline p_1^*$ and 
$p_2(t)\ra \overline p_2^*$, where $\overline p_1^*=\overline p_2^*=0.518$.
(b) $\al=0.1$. Both functions, $p_1(t)\ra p^*$ and $p_2(t)\ra p^*$, tend to the same 
limit, $p^*=1$, when $t\ra\infty$.
(c) $\al=0.3$. Both functions, $p_1(t)\ra p^*$ and $p_2(t)\ra p^*$, tend to the same 
limit, $p^*=1$, when $t\ra\infty$.
(d) $\al=0.5$. Both functions, $p_1(t)\ra p^*$ and $p_2(t)\ra p^*$, tend to the same 
limit, $p^*=1$, when $t\ra\infty$.

\vskip 1cm
{\bf Fig. 8}.
Probabilities $p_1(t)$ (solid line) and $p_2(t)$ (dashed line) for parameters 
$\ep_1=\ep_2=0$, $q_1=0.25$, $q_2=0.4$, $f_1=f_2=0.418$, $p^*=1$, and different $\al$:
(a) $\al=0$. For $t\ra\infty$, probabilities $p_1(t)\ra \overline p_1^*=f_1=0.418$ and 
$p_2(t)\ra \overline p_2^*=0.7408$.
(b) $\al=0.05$. Both functions, $p_1(t)\ra p^*$ and 
$p_2(t)\ra p^*$, tend to the same limit $p^*=1$, when $t\ra\infty$.
(c) $\al=0.2$. Both functions, $p_1(t)\ra p^*$ and $p_2(t)\ra p^*$, tend to the same 
limit, $p^*=1$.
(d) $\al=0.5$. Both functions, $p_1(t)\ra p^*$ and $p_2(t)\ra p^*$, tend to the same 
limit, $p^*=1$.

\vskip 1cm
{\bf Fig. 9}.
Probabilities $p_1(t)$ (solid line) and $p_2(t)$ (dashed line) for parameters 
$\ep_1=\ep_2=0$, $q_1=0.25$, $q_2=0.5819$, $f_1=f_2=0.418$, $p^*=1$, and different $\al$:
(a) $\al=0$. For $t\ra\infty$, probabilities $p_1(t)\ra \overline p_1^*=f_1=0.418$ and 
$p_2(t)\ra \overline p_2^*=0.8266$.
(b) $\al=0.1$. Both functions, $p_1(t)\ra p^*$ and 
$p_2(t)\ra p^*$, tend to the same limit $p^*=1$, when $t\ra\infty$.
(c) $\al=0.3$. Both functions, $p_1(t)\ra p^*$ and $p_2(t)\ra p^*$, tend to the same 
limit, $p^*=1$.
(d) $\al=1$. Both functions, $p_1(t)\ra p^*$ and $p_2(t)\ra p^*$, tend to the same 
limit, $p^*=1$.

\vskip 1cm
{\bf Fig. 10}.
The class of networks for the case of three groups of agents and three 
temporal slices.

\newpage

\begin{figure}[ht]
\centerline{
\hbox{
\includegraphics[width=8cm]{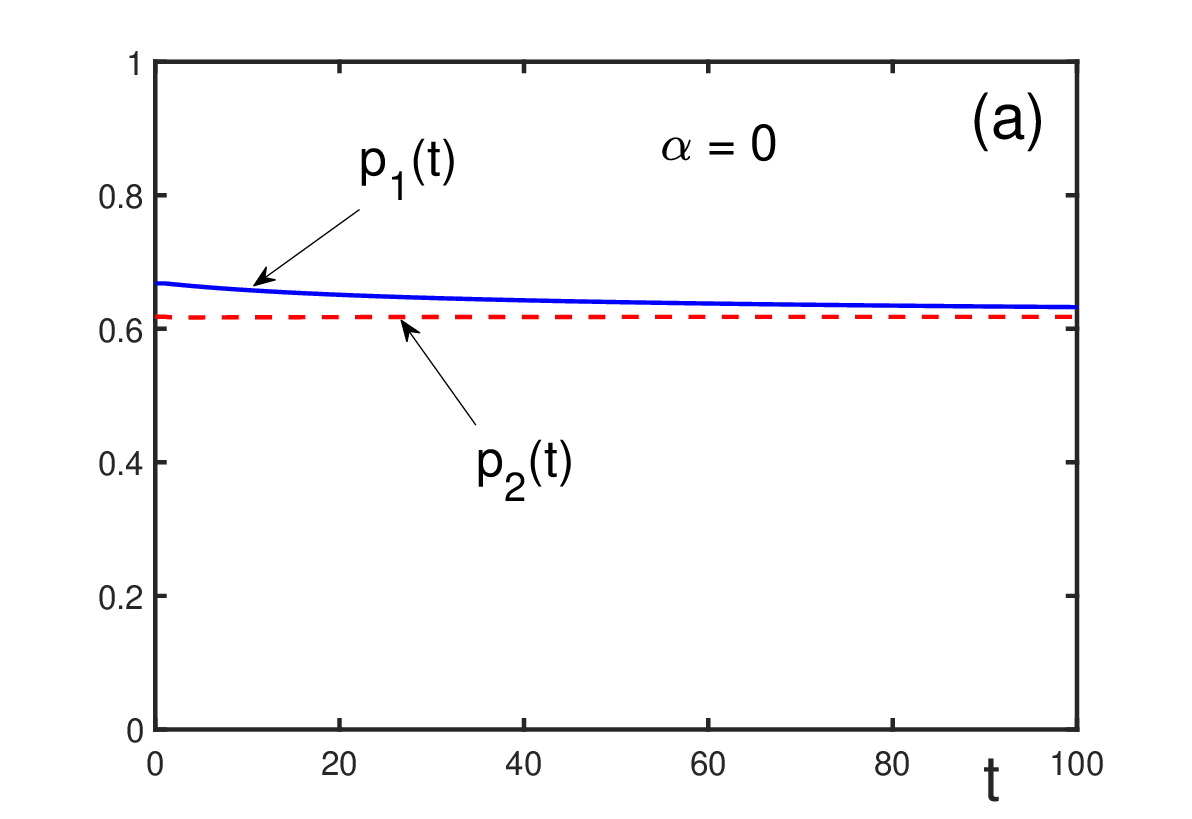} \hspace{1cm}
\includegraphics[width=8cm]{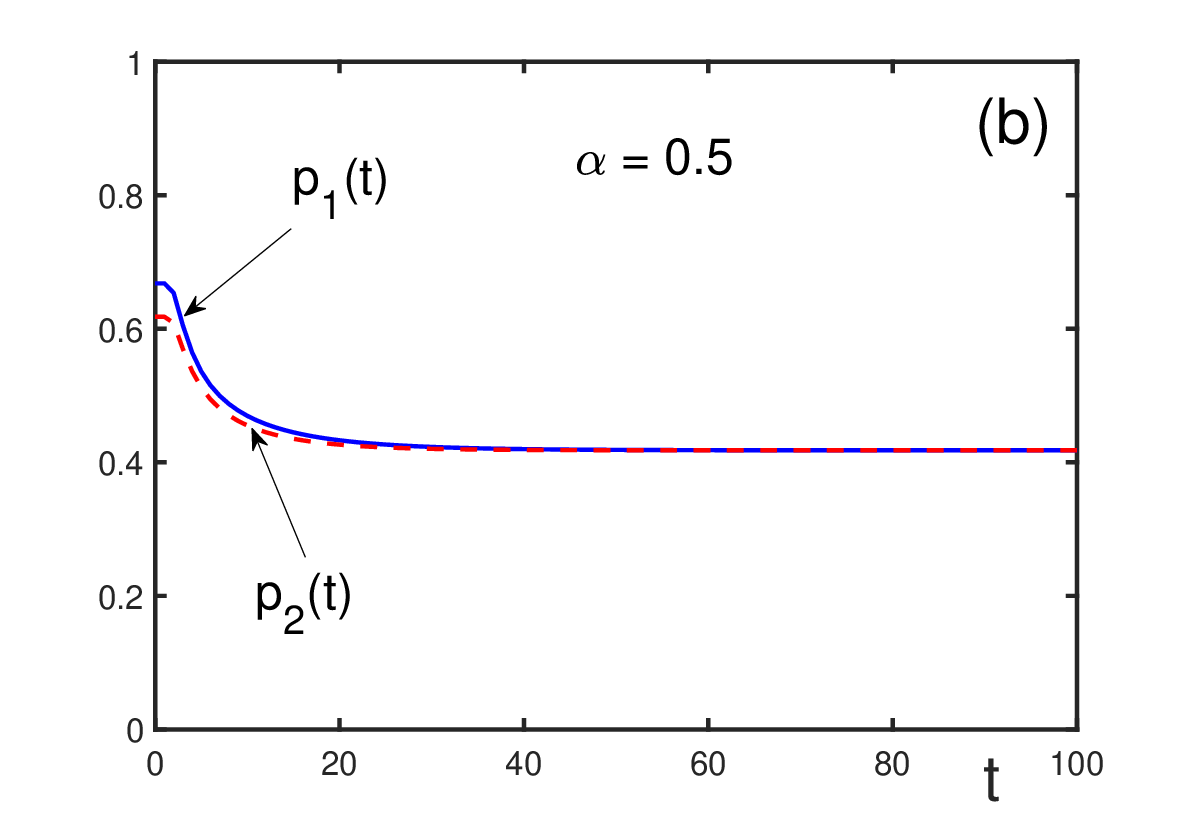} } }
\caption{\small
Probabilities $p_1(t)$ (solid line) and $p_2(t)$ (dashed line) for parameters 
$\ep_1=\ep_2=0$, $q_1=0.25$, $q_2=0.2$, $f_1=f_2=0.418$, $p^*=0,418$, and different $\al$:
(a) $\al=0$. For $t\ra\infty$, probabilities $p_1(t)\ra \overline p_1^*$ and 
$p_2(t)\ra \overline p_2^*$, where $\overline p_1^*=\overline p_2^*=0.618$.
(b) $\al=0.5$. Both functions, $p_1(t)\ra p^*$ and $p_2(t)\ra p^*$, tend to the same 
limit, $p^*=f_1=f_2=0.418$, when $t\ra\infty$.}
\label{fig:Fig.5}
\end{figure}

\begin{figure}[ht]
\centerline{
\hbox{
\includegraphics[width=8cm]{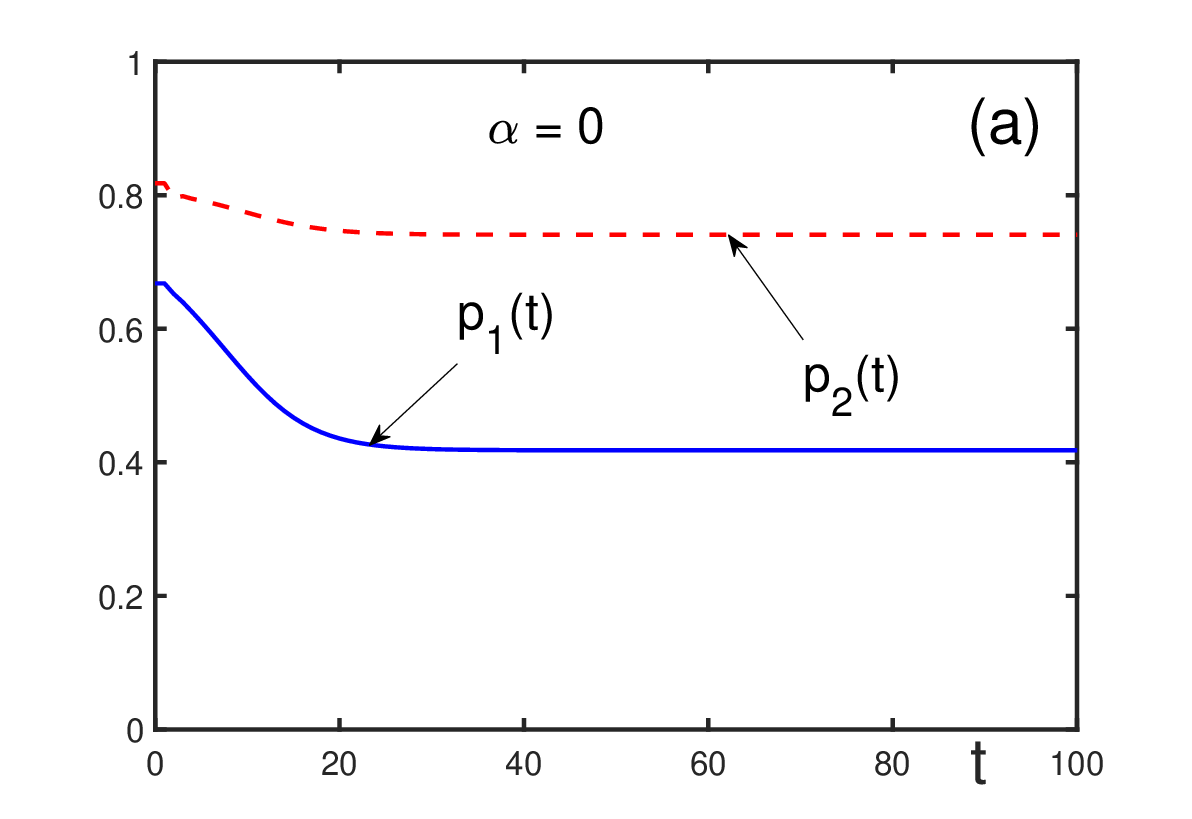} \hspace{1cm}
\includegraphics[width=8cm]{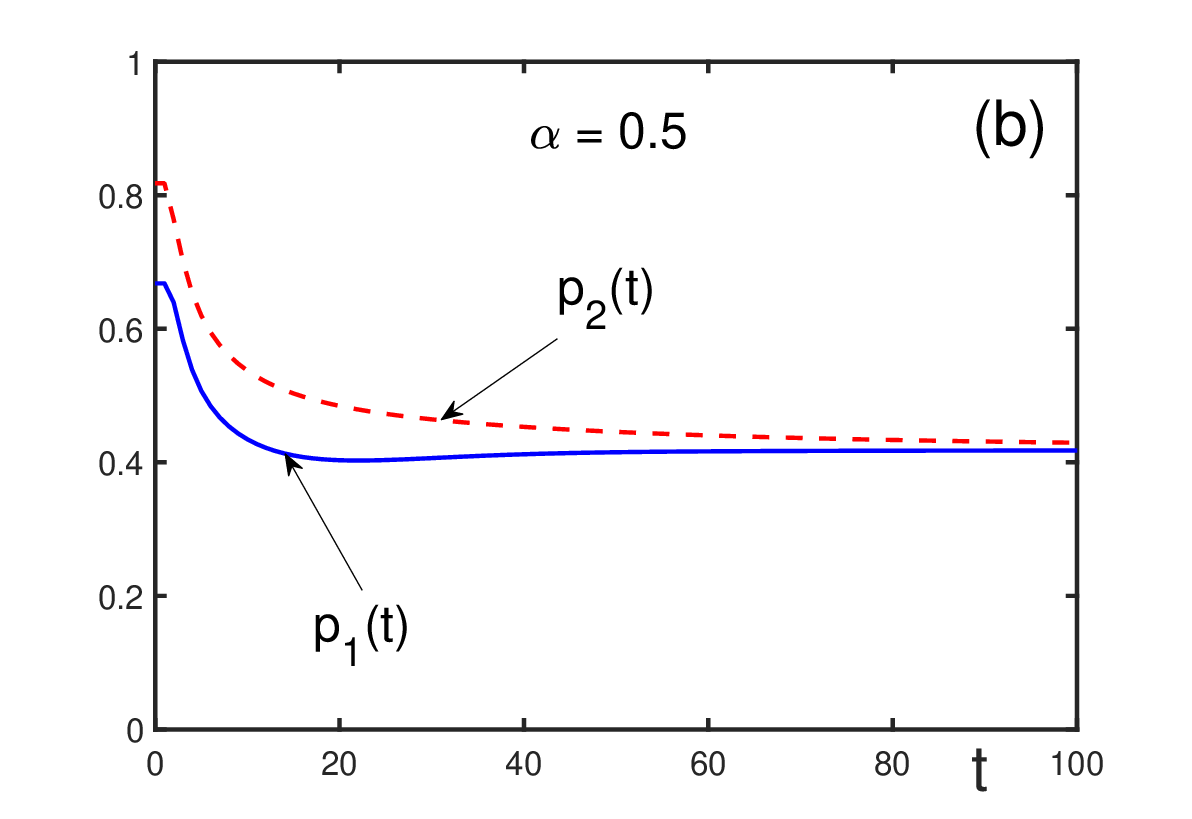} } }
\caption{\small
Probabilities $p_1(t)$ (solid line) and $p_2(t)$ (dashed line) for parameters 
$\ep_1=\ep_2=0$, $q_1=0.25$, $q_2=0.4$, $f_1=f_2=0.418$, $p^*=0,418$, and different $\al$:
(a) $\al=0$. For $t\ra\infty$, probabilities $p_1(t)\ra \overline p_1^*=p^*=f_1=0.418$ and 
$p_2(t)\ra \overline p_2^*=0.7408$.
(b) $\al=0.5$. Both functions, $p_1(t)\ra p^*$ and $p_2(t)\ra p^*$, tend to the same 
limit, $p^*=f_1=0.418$, when $t\ra\infty$.}
\label{fig:Fig.6}
\end{figure}

\begin{figure}[ht]
\centerline{
\hbox{
\includegraphics[width=8cm]{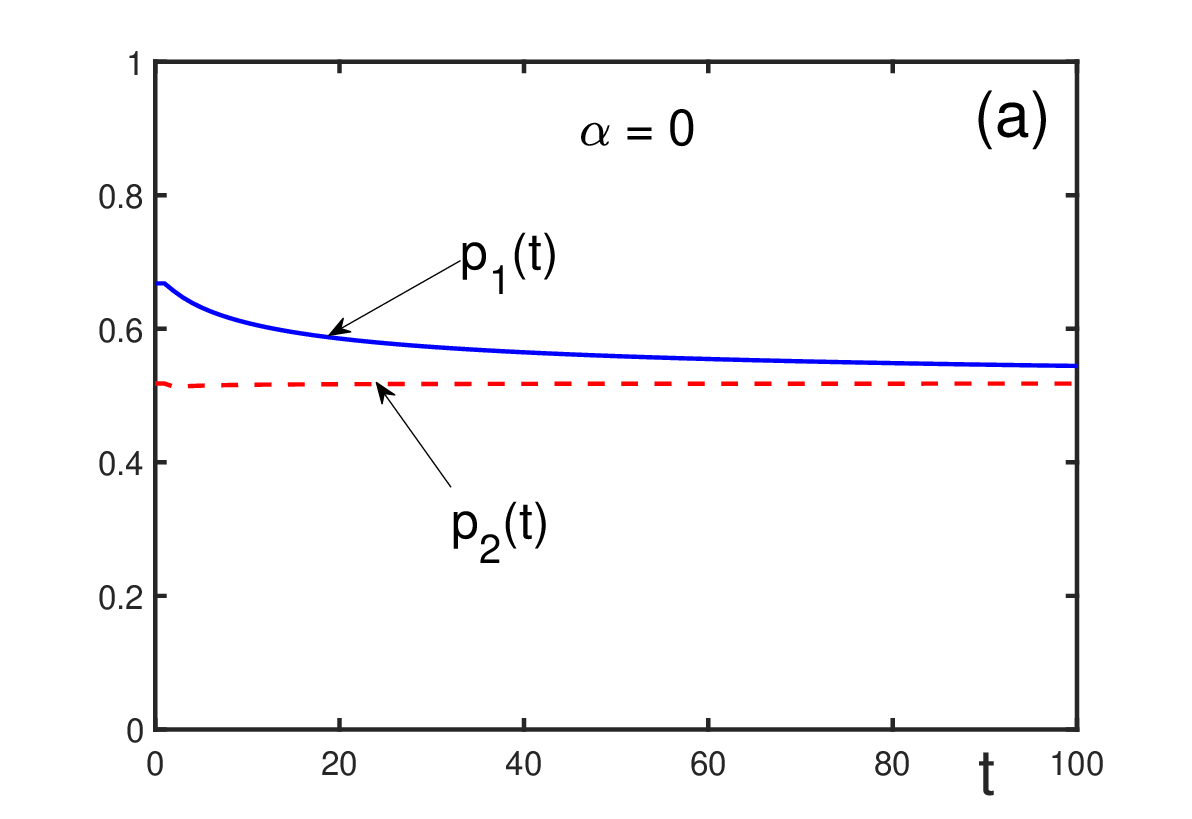} \hspace{1cm}
\includegraphics[width=8cm]{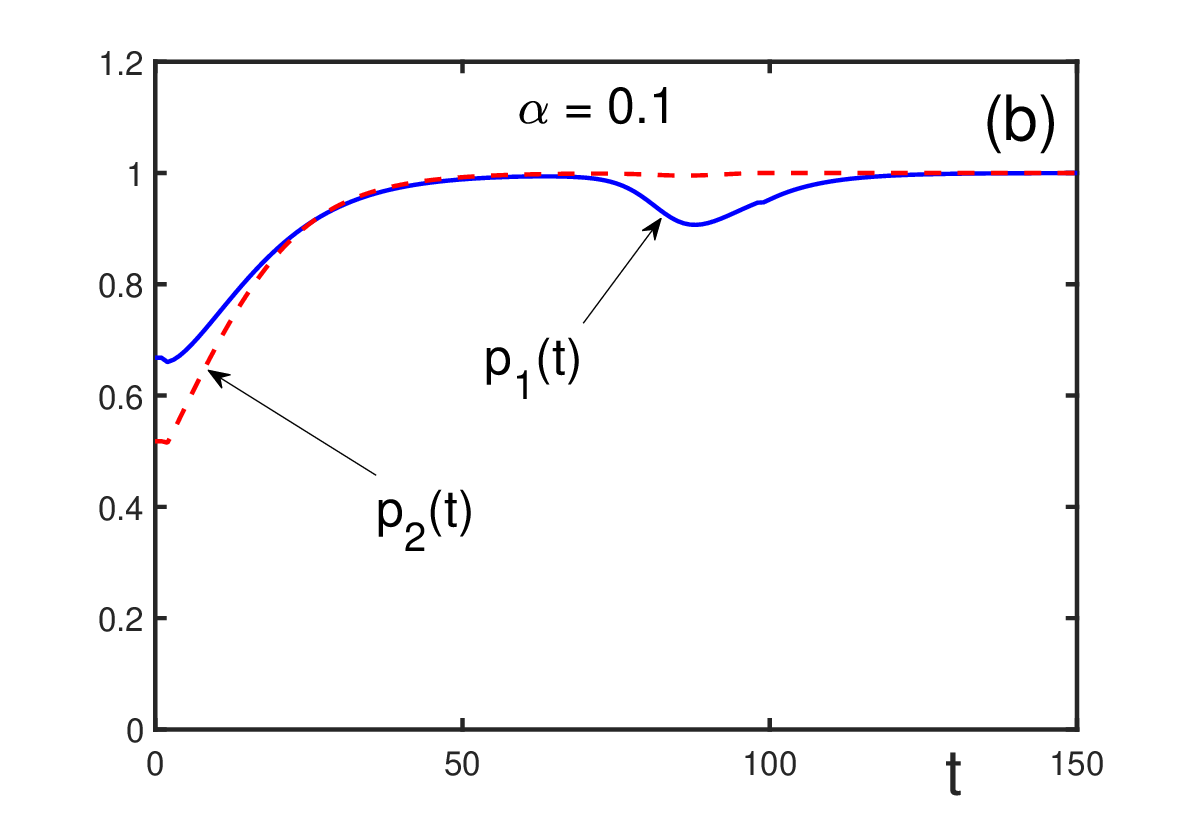} } }
\vskip 1cm
\centerline{
\hbox{
\includegraphics[width=8cm]{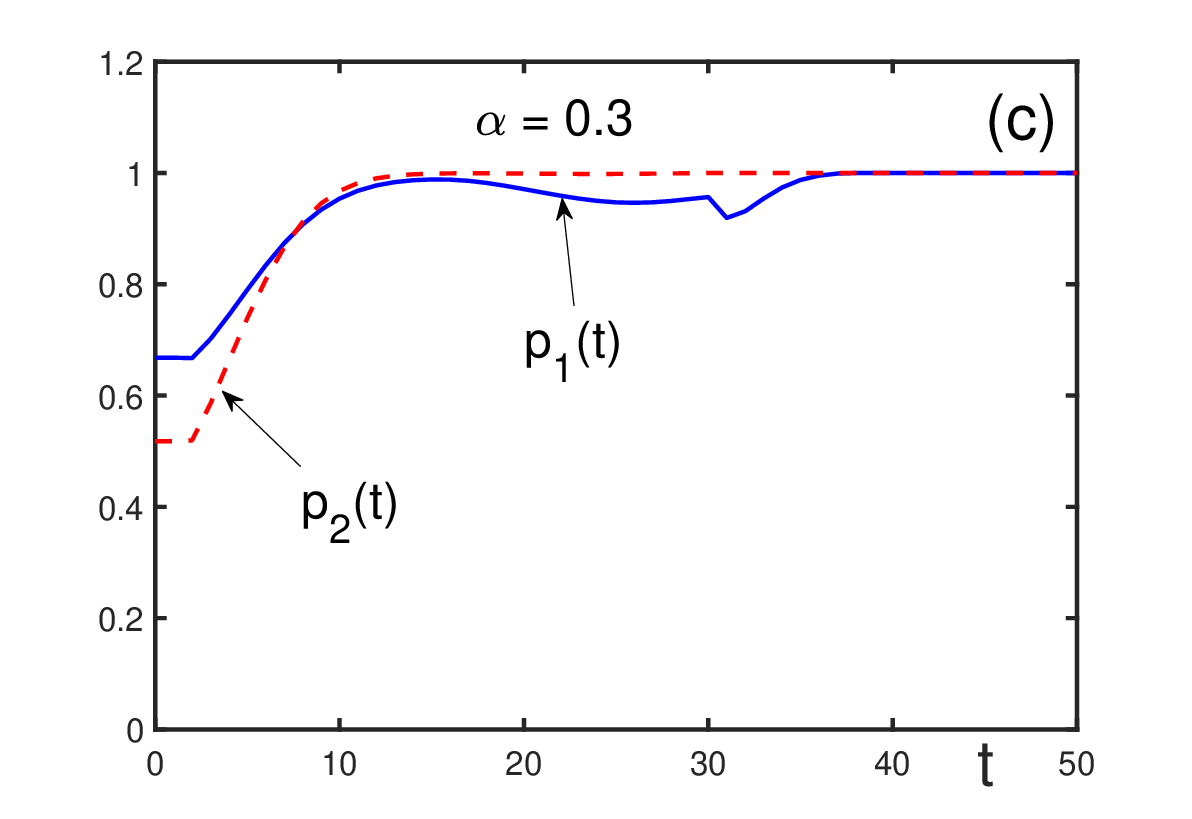} \hspace{1cm}
\includegraphics[width=8cm]{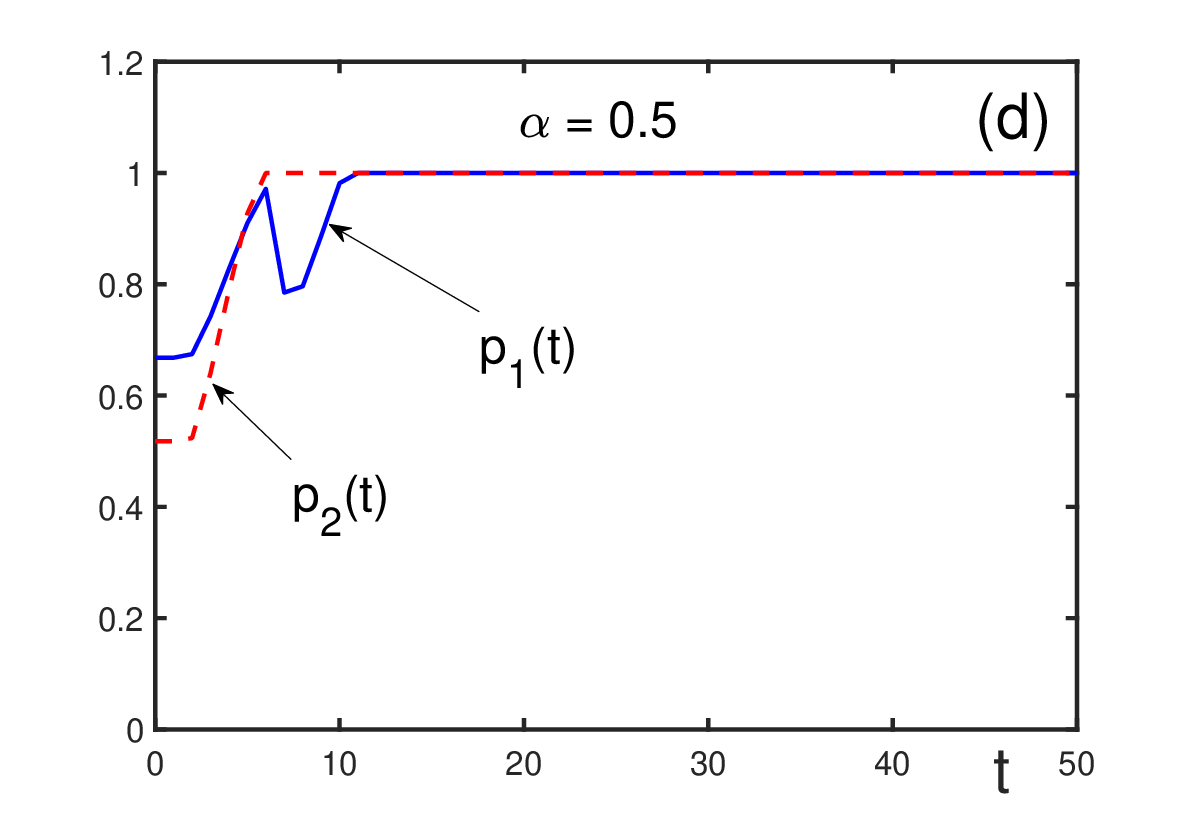} } }
\caption{\small
Probabilities $p_1(t)$ (solid line) and $p_2(t)$ (dashed line) for parameters 
$\ep_1=\ep_2=0$,  $q_1=0.25$, $q_2=0.1$, $f_1=f_2=0.418$, $p^*=1$, and different $\al$:
(a) $\al=0$. For $t\ra\infty$, probabilities $p_1(t)\ra \overline p_1^*$ and 
$p_2(t)\ra \overline p_2^*$, where $\overline p_1^*=\overline p_2^*=0.518$.
(b) $\al=0.1$. Both functions, $p_1(t)\ra p^*$ and $p_2(t)\ra p^*$, tend to the same 
limit, $p^*=1$, when $t\ra\infty$.
(c) $\al=0.3$. Both functions, $p_1(t)\ra p^*$ and $p_2(t)\ra p^*$, tend to the same 
limit, $p^*=1$, when $t\ra\infty$.
(d) $\al=0.5$. Both functions, $p_1(t)\ra p^*$ and $p_2(t)\ra p^*$, tend to the same 
limit, $p^*=1$, when $t\ra\infty$.
}
\label{fig:Fig.7}
\end{figure}

\begin{figure}[ht]
\centerline{
\hbox{
\includegraphics[width=8cm]{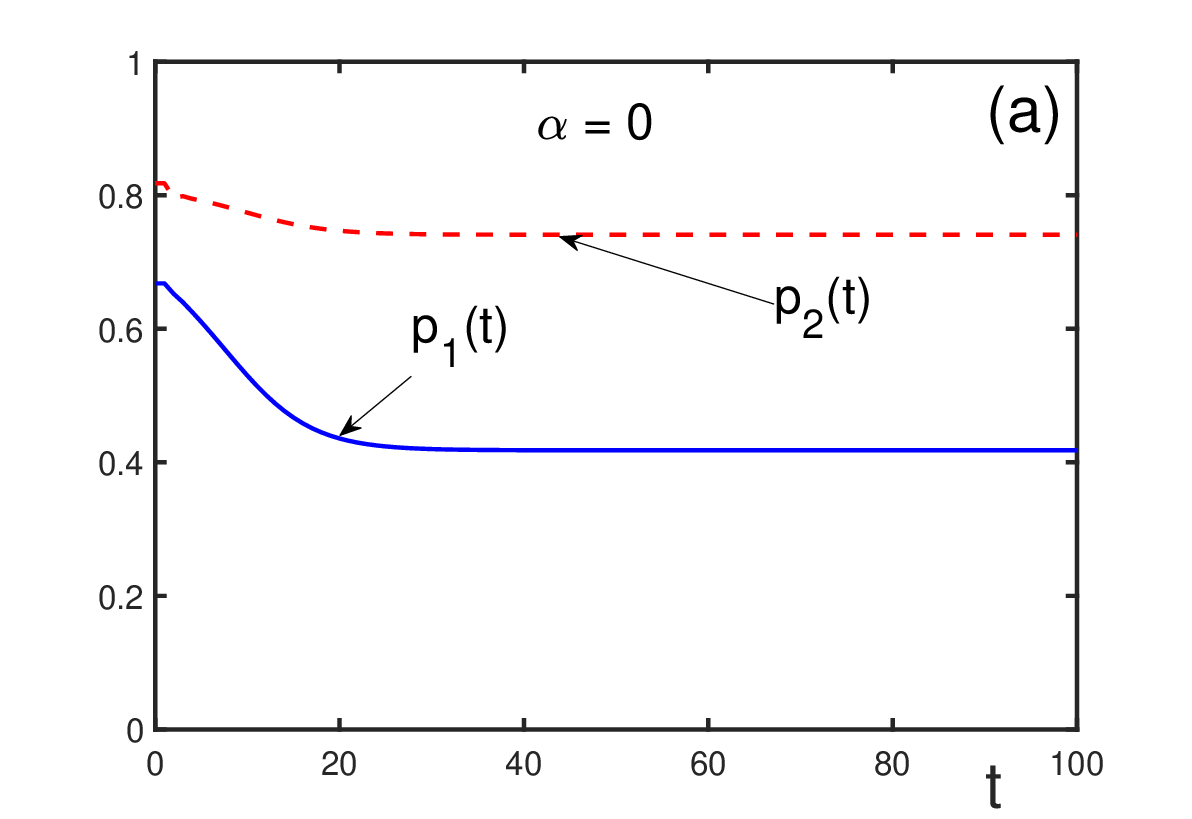} \hspace{1cm}
\includegraphics[width=8cm]{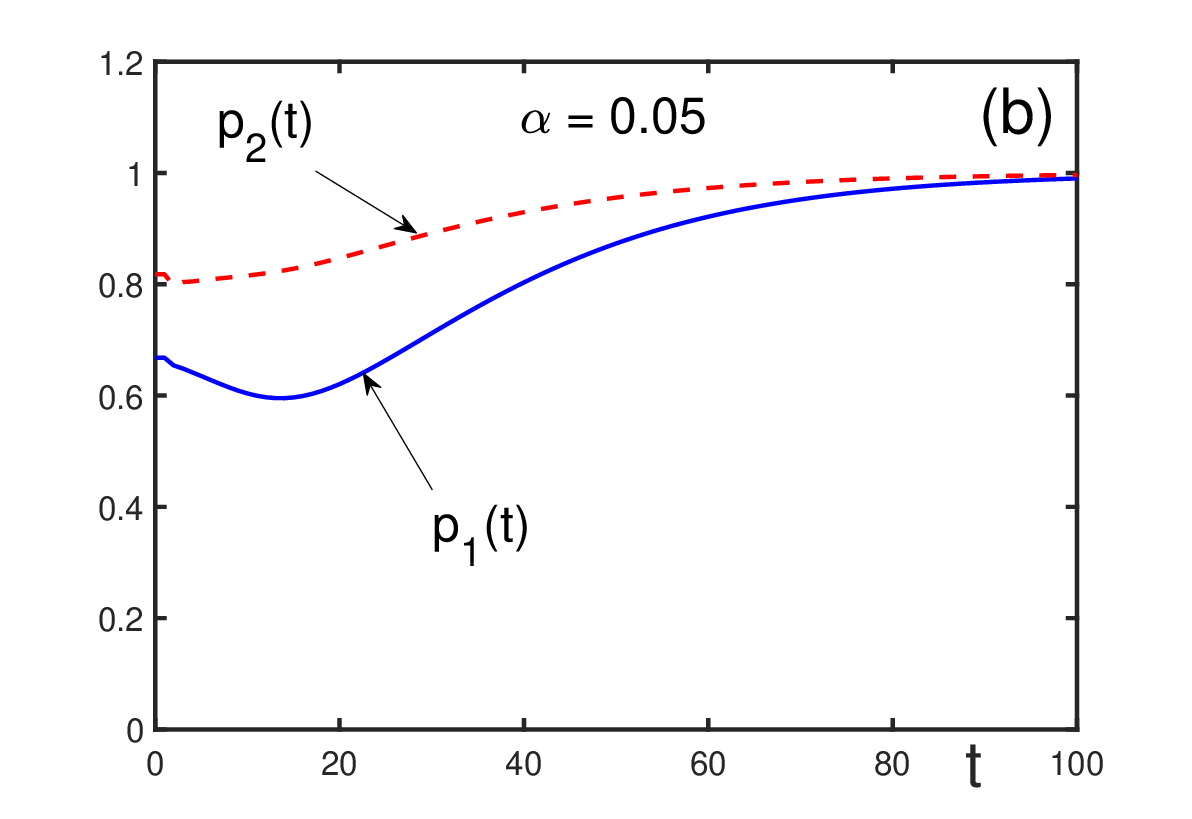} } }
\vskip 1cm
\centerline{
\hbox{
\includegraphics[width=8cm]{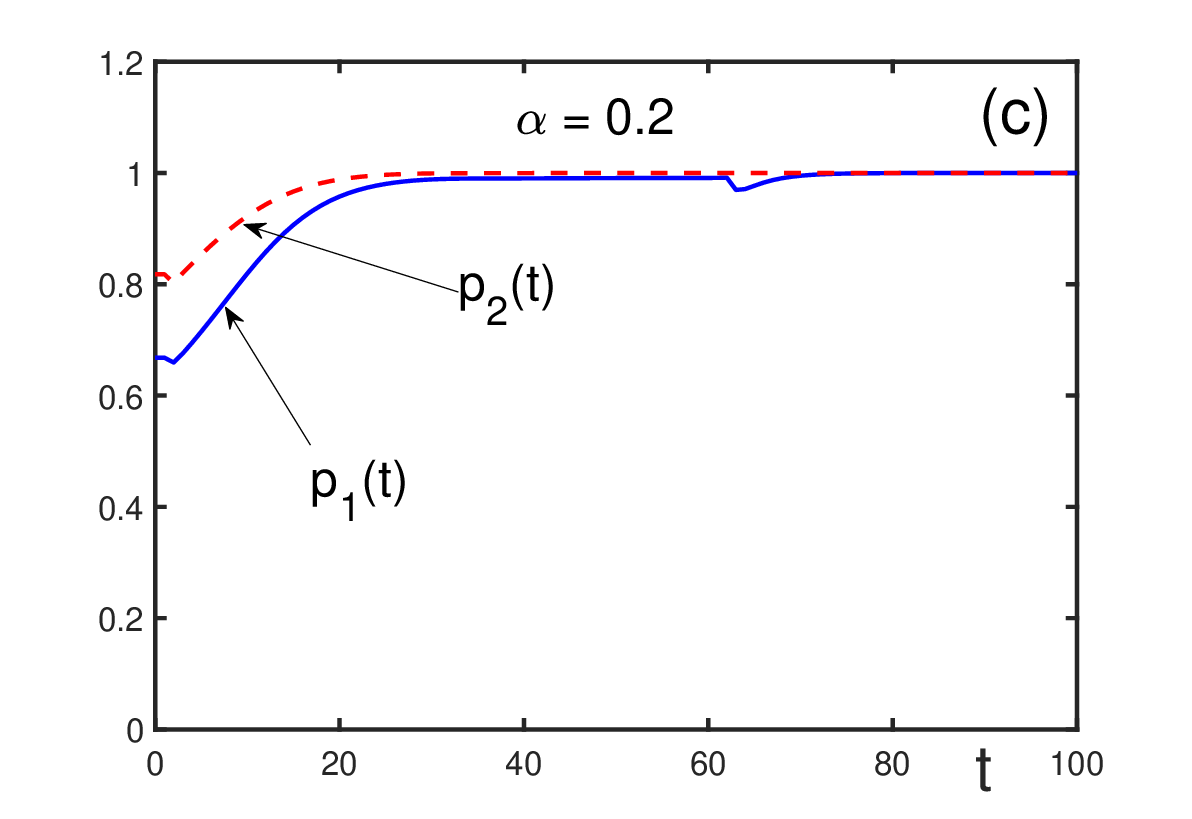} \hspace{1cm}
\includegraphics[width=8cm]{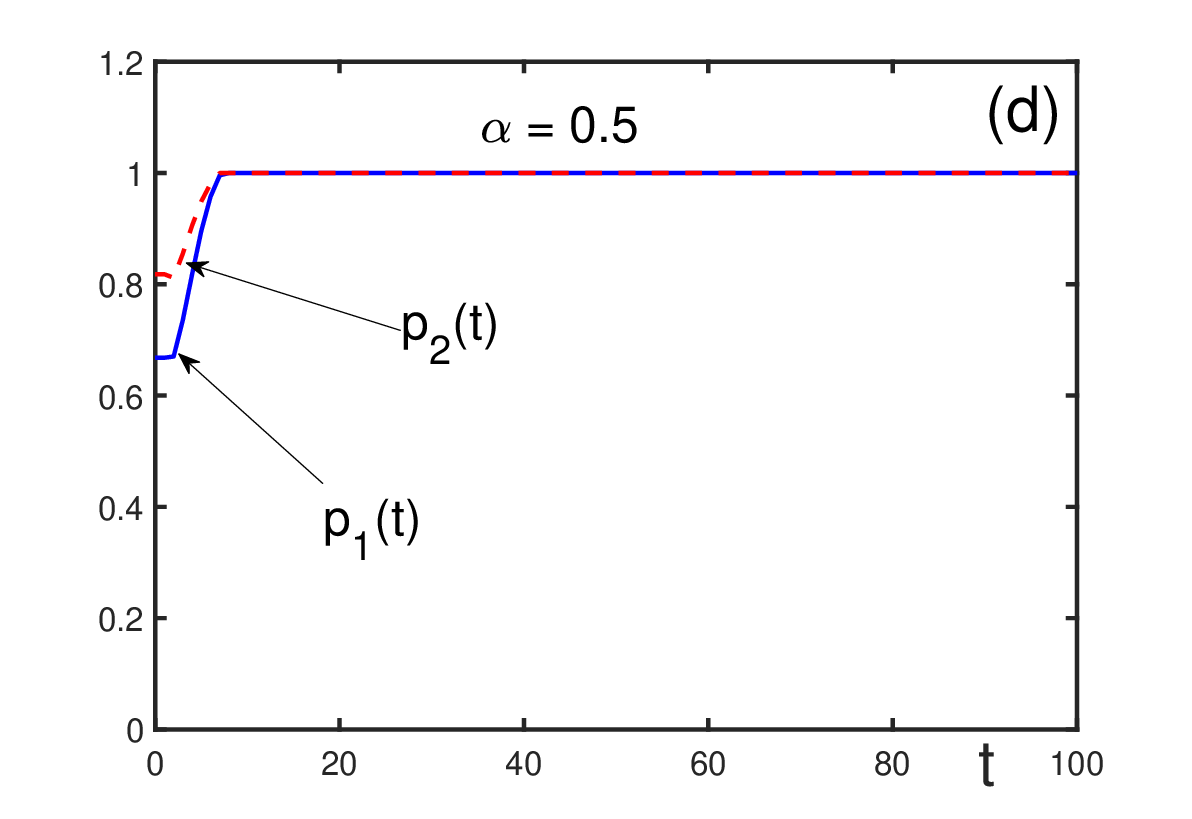} } }
\caption{\small
Probabilities $p_1(t)$ (solid line) and $p_2(t)$ (dashed line) for parameters 
$\ep_1=\ep_2=0$, $q_1=0.25$, $q_2=0.4$, $f_1=f_2=0.418$, $p^*=1$, and different $\al$:
(a) $\al=0$. For $t\ra\infty$, probabilities $p_1(t)\ra \overline p_1^*=f_1=0.418$ and 
$p_2(t)\ra \overline p_2^*=0.7408$.
(b) $\al=0.05$. Both functions, $p_1(t)\ra p^*$ and 
$p_2(t)\ra p^*$, tend to the same limit $p^*=1$, when $t\ra\infty$.
(c) $\al=0.2$. Both functions, $p_1(t)\ra p^*$ and $p_2(t)\ra p^*$, tend to the same 
limit, $p^*=1$.
(d) $\al=0.5$. Both functions, $p_1(t)\ra p^*$ and $p_2(t)\ra p^*$, tend to the same 
limit, $p^*=1$.
}
\label{fig:Fig.8}
\end{figure}

\begin{figure}[ht]
\centerline{
\hbox{
\includegraphics[width=8cm]{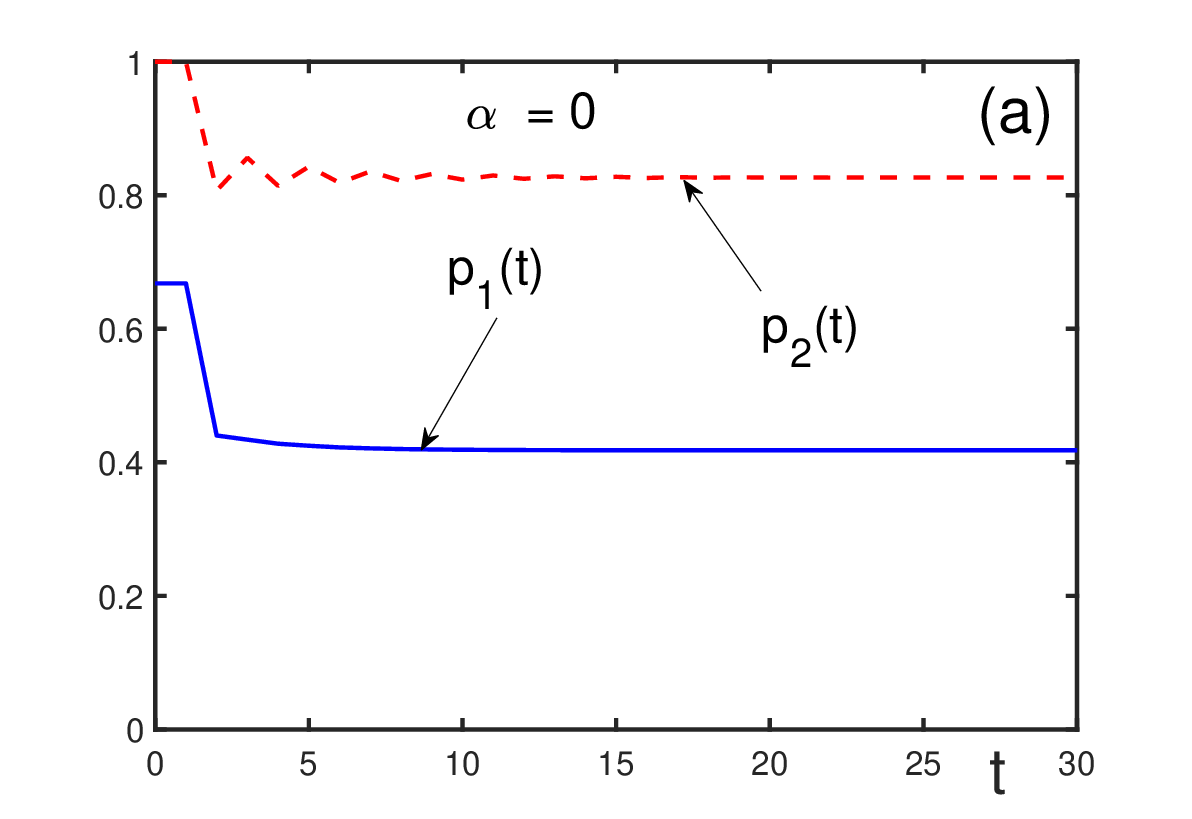} \hspace{1cm}
\includegraphics[width=8cm]{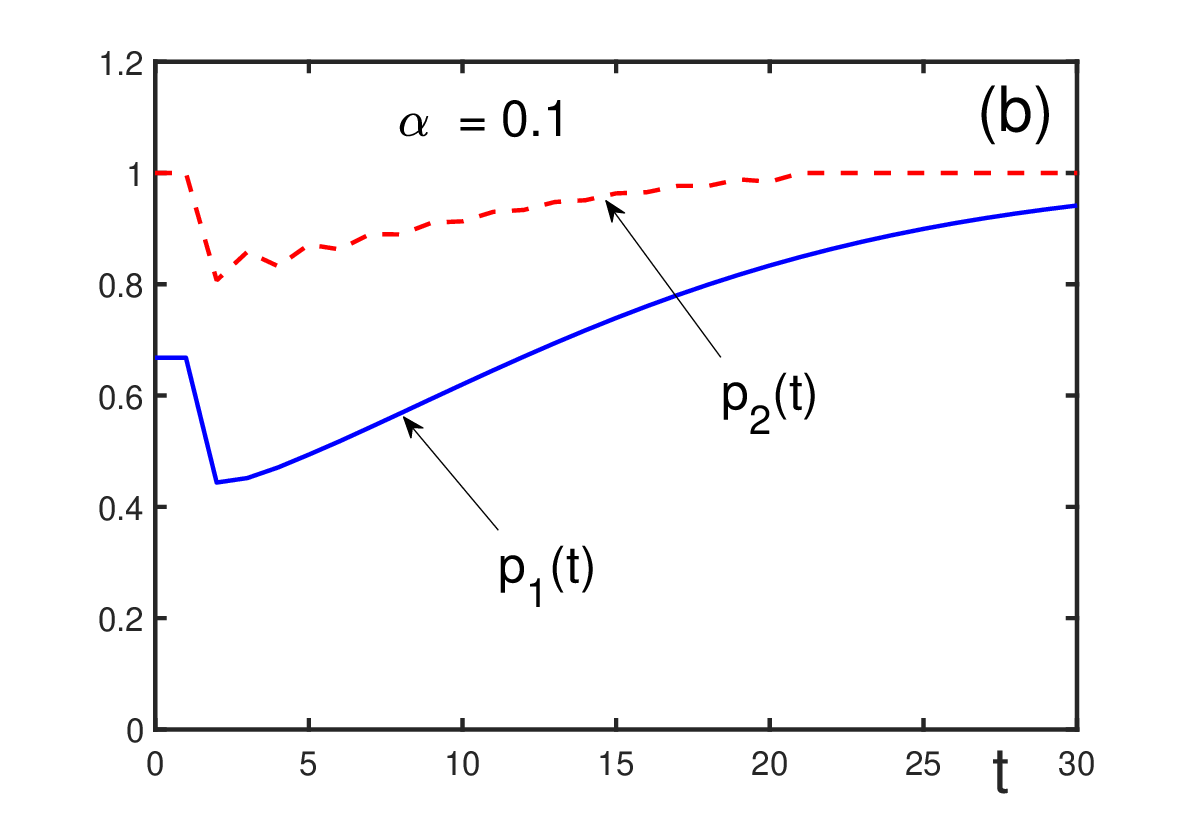} } }
\vskip 1cm
\centerline{
\hbox{
\includegraphics[width=8cm]{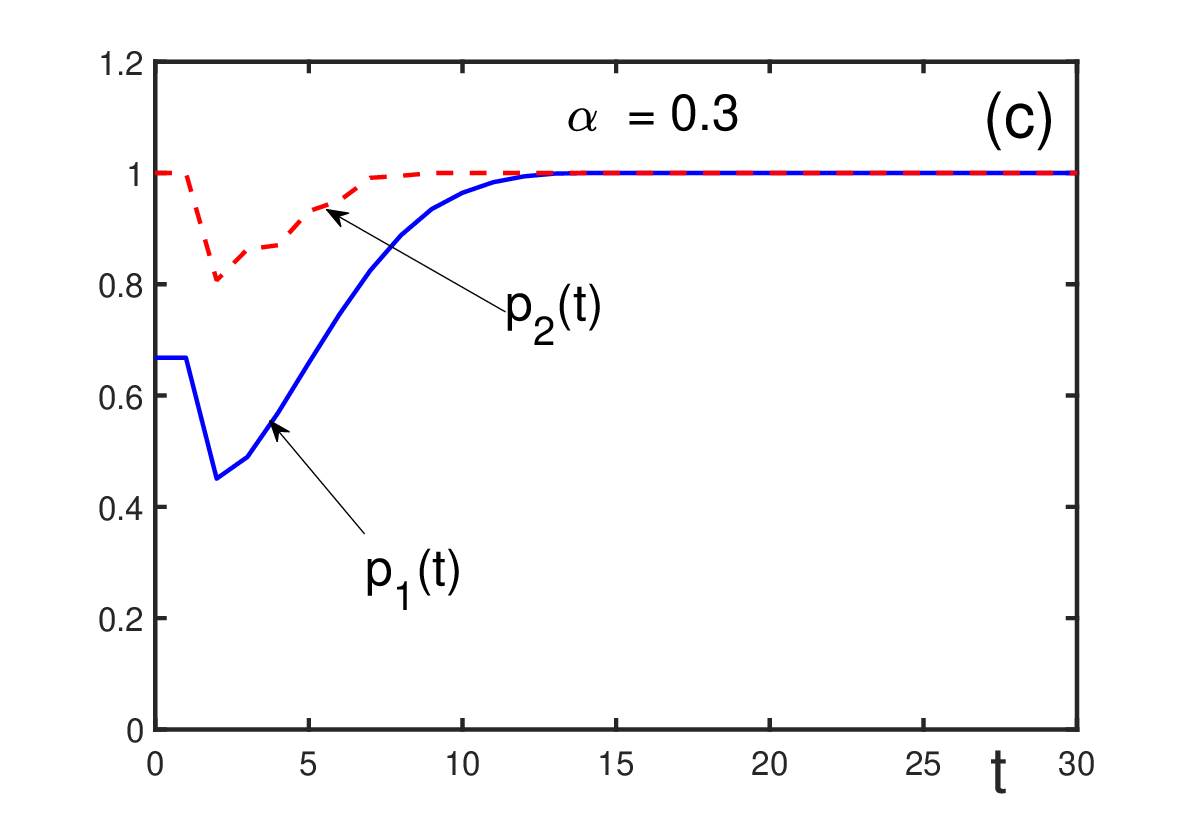} \hspace{1cm}
\includegraphics[width=8cm]{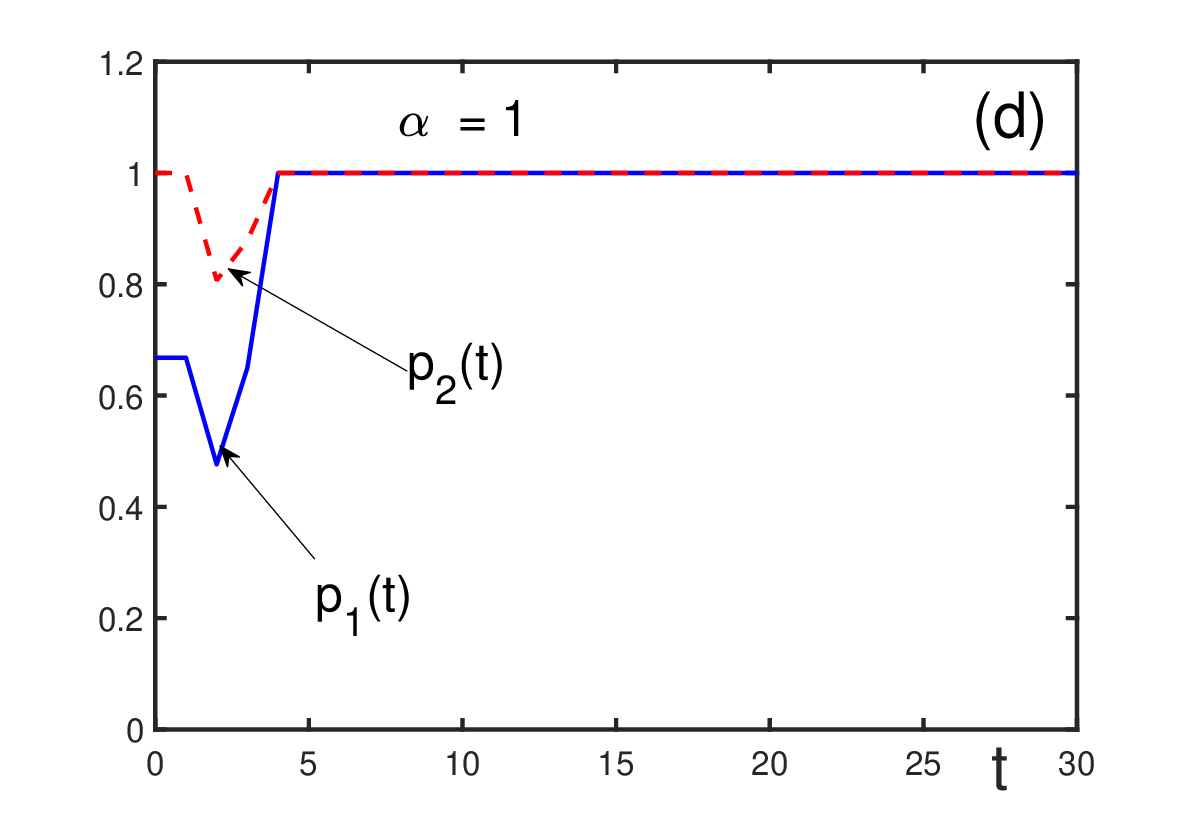} } }
\caption{\small
Probabilities $p_1(t)$ (solid line) and $p_2(t)$ (dashed line) for parameters 
$\ep_1=\ep_2=0$, $q_1=0.25$, $q_2=0.5819$, $f_1=f_2=0.418$, $p^*=1$, and different $\al$:
(a) $\al=0$. For $t\ra\infty$, probabilities $p_1(t)\ra \overline p_1^*=f_1=0.418$ and 
$p_2(t)\ra \overline p_2^*=0.8266$.
(b) $\al=0.1$. Both functions, $p_1(t)\ra p^*$ and 
$p_2(t)\ra p^*$, tend to the same limit $p^*=1$, when $t\ra\infty$.
(c) $\al=0.3$. Both functions, $p_1(t)\ra p^*$ and $p_2(t)\ra p^*$, tend to the same 
limit, $p^*=1$.
(d) $\al=1$. Both functions, $p_1(t)\ra p^*$ and $p_2(t)\ra p^*$, tend to the same 
limit, $p^*=1$.
}
\label{fig:Fig.9}
\end{figure}

\begin{figure}[ht]
\centerline{
\includegraphics[width=12cm]{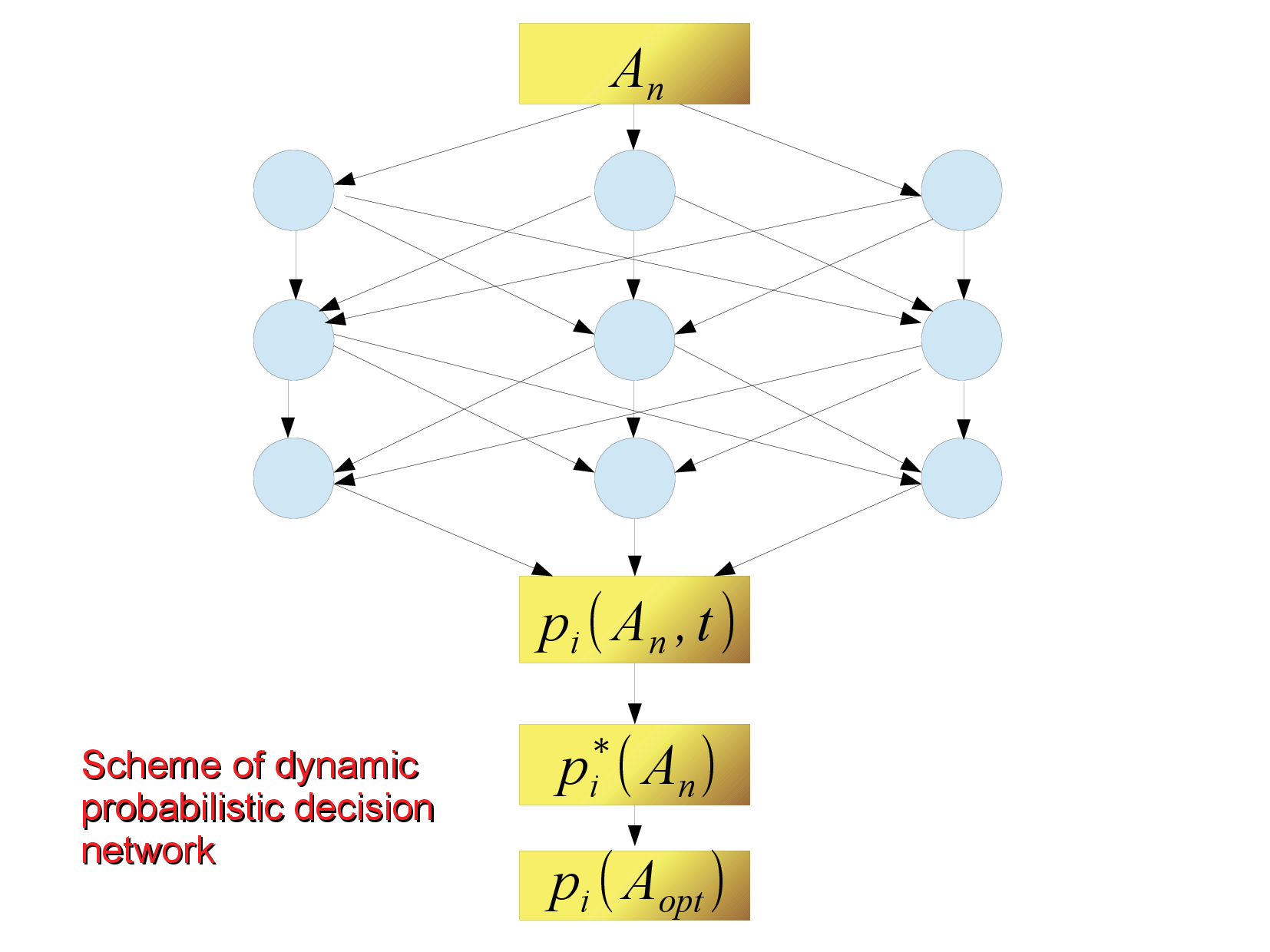} }
\caption{\small
The class of networks for the case of three groups of agents and three 
temporal slices.    
}
\label{fig:Fig.10}
\end{figure}

\end{document}